\documentclass[a4paper,11pt]{article}
\usepackage{jinstpub} 
\usepackage{lineno}

\usepackage{xcolor}
\usepackage{placeins}
\usepackage[normalem]{ulem}
\usepackage[caption=false]{subfig}

\newcommand{\tmu}{%
  \ifmmode
    \mathchoice
        {\hbox{\normalsize\textmu}}
        {\hbox{\normalsize\textmu}}
        {\hbox{\scriptsize\textmu}}
        {\hbox{\tiny\textmu}}%
  \else
    \textmu
  \fi
}

\usepackage{wrapfig}

\title{\center Time-Correlated Single-Photon Counting for\\ versatile longitudinal diagnostics \\at the MAX IV Laboratory storage rings}

\author[a,1]{Miriam Brosi,\note{Corresponding author.}}
\author[b]{Johann Schmand,}
\author[a]{Jonas Breunlin,}

\author[a,b]{and Francesca Curbis}
\affiliation[a]{MAX IV Laboratory, Lund University, Lund, Sweden}
\affiliation[b]{Department of Physics, Lund University, Lund, Sweden}

\emailAdd{miriam.brosi@maxiv.lu.se}

\abstract{Precise diagnostic on the electron beam parameters is a very valuable tool and essential in the operation of synchrotron light sources. One possible option is to employ the emitted synchrotron radiation for non-destructive measurements. A tool, which has been used in many ring based synchrotron light sources is Time-Correlated Single-Photon Counting (TCSPC). It allows to measure the arrival time distribution of the emitted photons and by that reveals the filling pattern, i.e., the charge distribution onto the electron bunches stored in the storage ring. At MAX IV, two TCSPC setups were installed and the analysis was developed further to also allow for the measurement of the longitudinal profiles of the individual bunches. The analysis is available as a Tango device in the accelerator control system and continuously provides, for example, the bunch length of each bunch as well as the bunch profiles and phases. This improved the diagnostic capabilities significantly, for example, in the presence of Landau cavities, which are becoming increasingly more common in new fourth-generation synchrotron light sources.}

\begin{document}
\maketitle
\flushbottom

\section{Introduction}

The longitudinal charge distribution is an important parameter in the operation of storage-ring-based synchrotron light sources. 
The so-called filling pattern describes the distribution of the electrons in the stored beam onto the multiple electron packages, i.e., the bunches.
Measurements of the filling pattern are in general used to, for example, ensure a high purity during single-bunch operation or the homogeneity of the distribution of charge onto the different bunches during multi-bunch operation.
Especially for light sources with passive higher harmonic cavities (also referred to as Landau cavities), such as the two storage rings of the MAX IV Laboratory, the distribution of the electrons onto the bunches is rather critical.
As the passive Landau cavities are driven by the electron beam itself, the filling pattern has a strong impact on the bunch-lengthening and the increase in synchrotron tune spread provided by the Landau cavities, and therefore, the stability of the bunches with respect to single- and multi-bunch collective effects.

Time-correlated single-photon counting (TCSPC) is commonly used as a non-invasive method to determine the filling pattern via the emitted synchrotron light in storage ring synchrotron light sources. 
New TCSPC setups have been commissioned for each of the two storage rings at the MAX IV Laboratory.
The first half of this paper discusses the new setups and the relevant properties of the detectors and devices used as well as describes shortly the integration into the accelerator control system. 
Additionally, an overview of different use cases for the filling pattern measurement is given.  

The second half of this paper focuses on a newly developed method to measure the longitudinal bunch profiles. 
It was discussed in the past that, given a high enough temporal resolution, it should be possible to resolve not only the distribution of the electrons onto the different bunches but also the distribution of the electrons within each bunch~\cite{yang_bunch-by-bunch_2010}.
The main challenge was the influences of detector and timing effects on the measured distributions. 
Here, we present a new method developed within the framework of a thesis project at MAX IV~\cite{schmand_johann_measuring_2024}.
The method corrects for these influences and therefore allows the measurement of the individual longitudinal bunch profiles.
The system's transit-time spread (TTS) was experimentally determined and an algorithm was developed to extract the original charge distribution from the measured histogram via deconvolution. Based on the deconvolved bunch profiles further properties can be calculated, such as the bunch length and relative phase of each bunch.
The method was successfully tested and several examples for its application are presented in the following sections.

\section{Time-Correlated Single-Photon Counting}

Time-correlated single-photon counting is a method used, for example, in time-resolved fluorescence spectroscopy to record time-dependent intensity profiles of fluorescence light emitted after excitation (e.g.~\cite{wahl_time-correlated_2014}).
The principle behind TCSPC can be described rather simply. Single photons are detected and their arrival time is measured with respect to a reference trigger signal. By collecting such events over several cycles, the resulting histogram of arrival times reveals the statistical distribution of the observed photon emission process.

\begin{figure}
\centering
\includegraphics[width=0.99\textwidth]{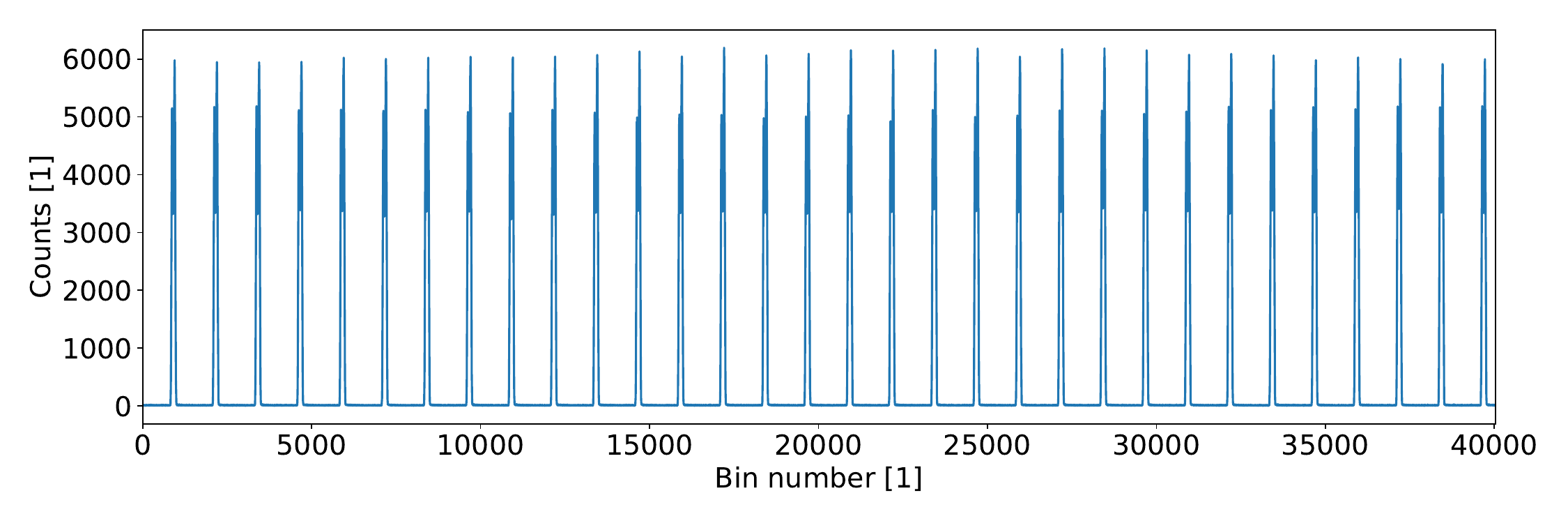}
\vspace{-0.5cm}
\caption{Example measurement of time-correlated single-photon counting (TCSPC) at the 1.5\,GeV storage ring. The histogram shows the correlation between arrival time and number of detected synchrotron light photons from the 32 stored electron bunches.  \label{fig:histogramR1}}
\end{figure}

This possibility to make the arrival-time distribution of events visible, makes TCSPC a perfect tool for measuring the charge distribution of electrons within the electron beam of a synchrotron light source storage ring, often referred to as filling pattern.
The emitted number of photons in incoherent synchrotron radiation is proportional to the number of stored electrons and shows the same temporal distribution.
Therefore, a TCSPC measurement of the emitted synchrotron light can be used as non-destructive diagnostic tool to determine the temporal charge distribution within the stored electron beam.
While this could also be done with fast photo-diode detectors and oscilloscopes, it would require the detector signal to be proportional to the number of photons emitted.
A significantly higher amplitude resolution can be achieved with TCSPC measurements.
The synchrotron radiation, typically in the visible range, is attenuated strongly and detected by a single-photon detector.
The detected photon events are then counted with a histogramming device and sorted according to arrival time relative to an external trigger.
For the application of measuring the filling pattern in a storage ring, this reference trigger is the revolution clock, triggering once per revolution around the storage ring.
The resulting histogram gives a correlation between arrival time and number of detected photons as shown in figure~\ref{fig:histogramR1}. 
The observed clusters correspond to the electron bunches stored in the storage ring with empty areas in between. 
Differences in the height of the clusters indicate the relative distribution of the charge onto the stored electron bunches.

TCSPC setups, for the purpose of filling pattern detection, are and have been used at many synchrotron light sources around the world, among others at  ALBA~\cite{torino_filling_2014, torino_time_2017}, DIAMOND~\cite{thomas_bunch_2006, thomas_time_2007, rehm_different_2014}, BESSY~II~\cite{holldack_bunch_2007}¸ Taiwan Light Source~\cite{wu_filling_2008}, SPEAR 3~\cite{corbett_bunch_2014, cope_upgrades_2016, xu_electron_2018}, KARA~\cite{kehrer_visible_2015, kehrer_filling_2018}, APS~\cite{yang_bunch-by-bunch_2010}. While in most cases the visible range of the synchrotron radiation is used, the setups can vary. Different detector types have been tested \cite{rehm_different_2014} for their suitability in the past. In the following section, the setups chosen for the two storage rings at the MAX IV Laboratory will be described.

\subsection{Setup at MAX IV}

\begin{figure}
\centering
\includegraphics[width=0.49\textwidth]{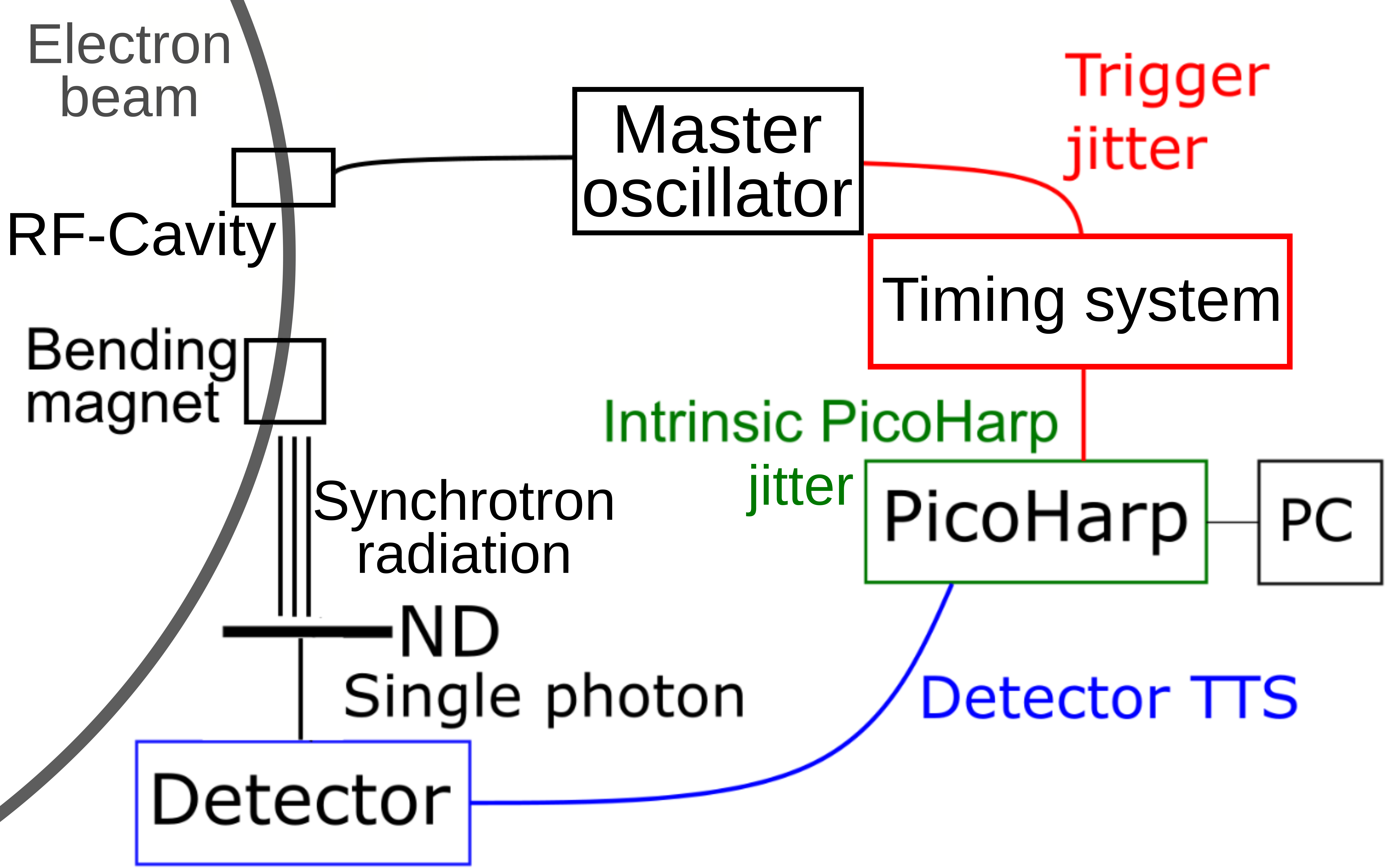}%
  \qquad%
\includegraphics[trim=0cm 0cm 15cm 15cm, clip, width=.45\textwidth]{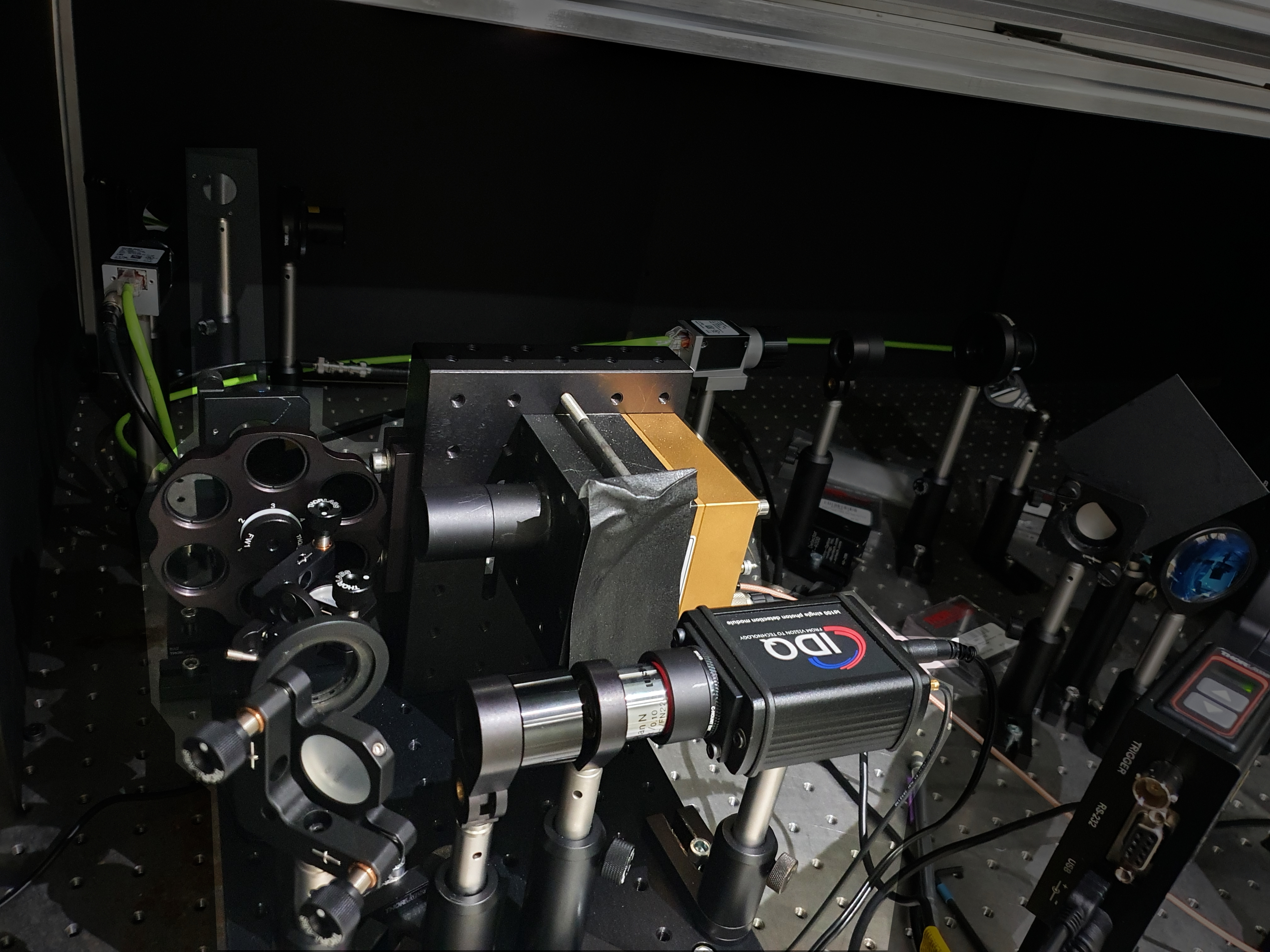}
\caption{TCSPC setup shown as sketch (left) including all the timing and jitter contributions and a photo of the optical setup (right) at the 1.5\,GeV ring visible light diagnostic beamline showing a filter wheel with neutral density (ND) filters and the PMA and the SPAD detectors.  \label{fig:setup}}
\end{figure}

A TCSPC setup (figure~\ref{fig:setup}) has been built at both storage rings at the MAX IV Laboratory. The setups are situated at the visible light diagnostic beamlines~\cite{breunlin_emittance_2016}.
In both cases the setup consists of a single-photon detector and a PicoHarp from PicoQuant~\cite{picoquant_notitle_nodate}
as the histogramming device.
Additionally, neutral density filters are used to strongly attenuate the synchrotron radiation light to allow the use of single-photon detectors. The sustainable count rate depends on the specific detector as well as on the  histogramming device used.

\subsubsection*{Single-Photon detectors}
\begin{figure}
\centering
\includegraphics[width=0.98\textwidth]{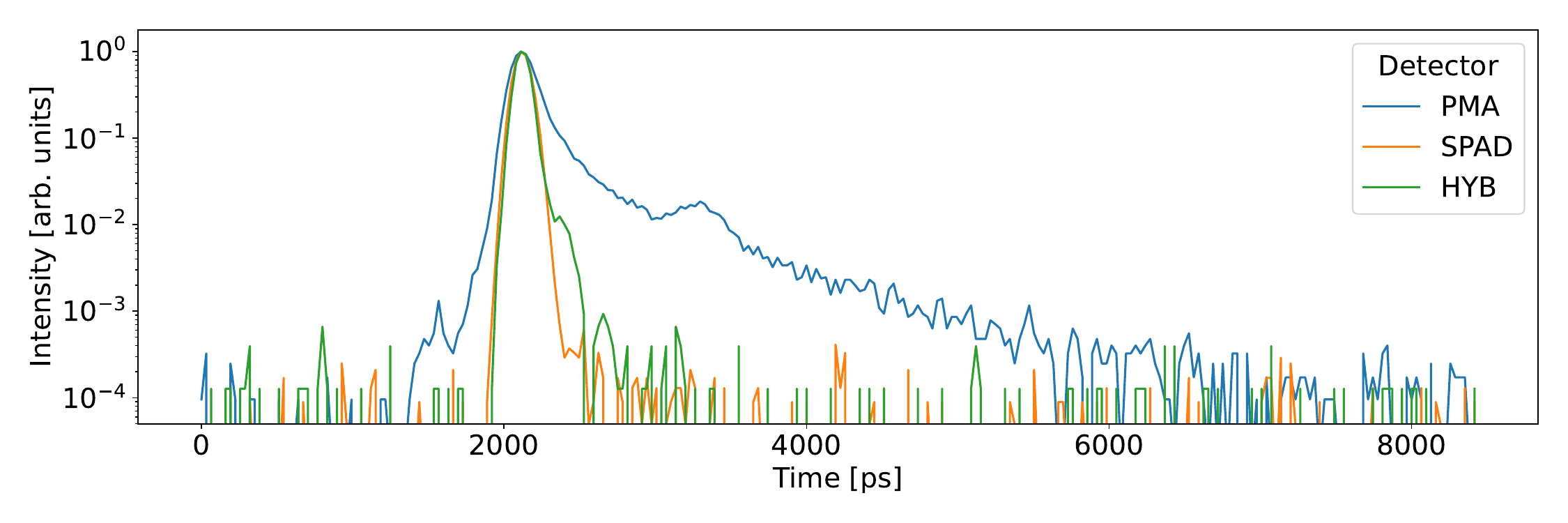}
\vspace{-0.5cm}
\caption{Raw histograms of a short (39\,ps rms) Gaussian bunch measured at the 3\,GeV ring showing the differences in the transit-time spreads caused by the three different detector types Photomultiplier Assembly (PMA), Single-Photon Avalanche Diode (SPAD), and Hybrid-Photomultiplier Assembly (HYB). \label{fig:detector_comp}}
\end{figure}

The following three detector types are used as single-photon detectors in the TCSPC setups at the MAX IV diagnostic beamlines.
\subparagraph{Photomultiplier Assembly}
The PMA 175 (Photomultiplier Assembly) from PicoQuant~\cite{picoquant_pma} consists of a photo-multiplier tube and has a sensitive area with 8\,mm diameter. The wide detector diameter makes this detector insensitive to smaller movements of the transverse beam position caused by, for example, orbit corrections or bunch excitations for betatron tune measurements.
Disadvantages of this detector are the dark count rate of approximately 50\,counts/s and the fluctuation of the detector pulse amplitude common to photo-multiplier detectors.
Another downside is that the transit-time spread function of the detector is not a smooth peak but shows a significant after-pulse visible as a "shoulder" after the main peak in the measured histogram, with the main peak having a FWHM of $>90$\,ps.
Figure~\ref{fig:detector_comp} shows a measured histogram of a short, Gaussian electron bunch (with a known bunch length of 92\,ps FWHM) in the 3\,GeV ring. The logarithmic scale clearly shows the after-pulsing of the PMA detector. 

\begin{wrapfigure}{R}{0.5\textwidth}
\centering
\includegraphics[width=0.49\textwidth]{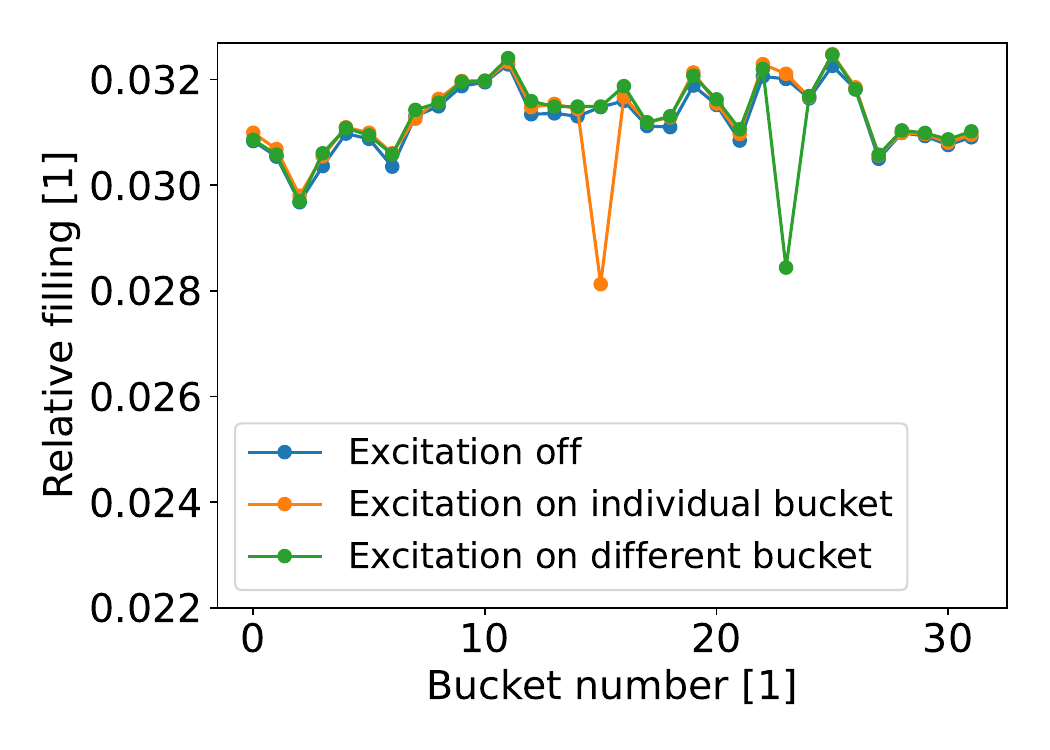}
\vspace{-0.5cm}
\caption{\label{fig:tuneexcitation} Filling patterns measured with the SPAD detector (without additional focusing onto the detector) without (blue) and with (orange and green) tune excitation in different individual bunches show a temporary and reversible reduction by about 10\% for the excited bunch. }
\end{wrapfigure}

\subparagraph{Single-Photon Avalanche Diode}
The ID100 single-photon avalanche diode (SPAD) from IDQuantique~\cite{idquantique_id100} comes with different sensitive area sizes and noise levels. 
We use the ID100-20-ULN version with a sensitive area of 20\,\tmu m diameter due to its low dark count rate of less than 7\,counts/s.
The spectral sensitivity ranges from 350\,nm to 900\,nm. 
While the SPAD shows significantly less after-pulsing than the PMA detector, it suffers from the "diffusion tail" effect caused by the generation of electron-hole pairs in the neutral layer below the depletion layer in the diode element~\cite{tosi_fast-gated_2011}.  
The effect is wavelength-dependent, which allows us to suppress the diffusion tail by adding a 405\,nm bandpass filter in front of the SPAD. 
The resulting overall shape of the TTS forms a nearly symmetric  pulse~\cite{becker_tcspc_2005} with a FWHM width in the order of 60\,ps.
The symmetric shape is very well visible in the comparison with the other detectors in figure~\ref{fig:detector_comp}.
This makes the usage of the SPAD more suitable than the PMA detector for the determination of the bunch profile from the detected TCSPC histogram. 
However, the very small sensitive area makes both, aligning the detector to the beam and maintaining a consistent count rate, more difficult. 
For example, an excitation applied to one of the electron bunches, in order to measure the betatron tunes, affects the measured filling pattern significantly, as can be seen in figure~\ref{fig:tuneexcitation}.
The small blowup in transverse size results in a decrease of about 10\% for the example measurement in the number of counted photons from the excited bunch versus the number of photons counted for the unexcited bunches. 
This reduction is reversed as soon as the excitation is switched off.
This effect could be reduced to a few percent by adding additional focusing elements in the light path, but still remains visible. While the detector is available with larger sensitive areas, this comes with the downside of increased dark count rates.

\subparagraph{Hybrid-Photomultiplier Assembly}
The hybrid-photomultiplier assembly (PMA Hybrid 06) detector from PicoQuant~\cite{picoquant_pma_hybrid} combines the bigger sensitive area of the PMA with the low dark count rate and shorter TTS of the SPAD. 
The sensitive area is 6\,mm in diameter, making it insensitive to bunch excitations and smaller movements of the beam position. 
The dark count rate is low, with less than 10\,counts/s, resulting in a good signal-to-noise ratio and therefore, e.g., a high measurable single-bunch purity.
While the PMA Hybrid still has some after-pulsing in the TTS, it is by a factor 2 lower than in the PMA and closer to the main pulse. Furthermore, the FWHM of the transit-time-spread is shorter at approximately 50\,ps and comparable to the SPAD (see figure~\ref{fig:detector_comp}).
This combination of properties makes the PMA Hybrid the best suited of the three detector types for measurements of the filling pattern and longitudinal bunch profiles via TCSPC.

Despite the described differences, all three detector types produce good results for TCSPC measurements of the longitudinal charge distribution, with the PMA and the PMA Hybrid being more stable for filling pattern measurements and the SPAD and the PMA Hybrid being more suitable for determining the longitudinal bunch profiles. At the moment, all three detectors are used distributed over the two storage rings at MAX IV.

\subsubsection*{Histogramming devices}
\begin{wrapfigure}{R}{0.5\textwidth}
\centering
\includegraphics[width=0.49\textwidth]{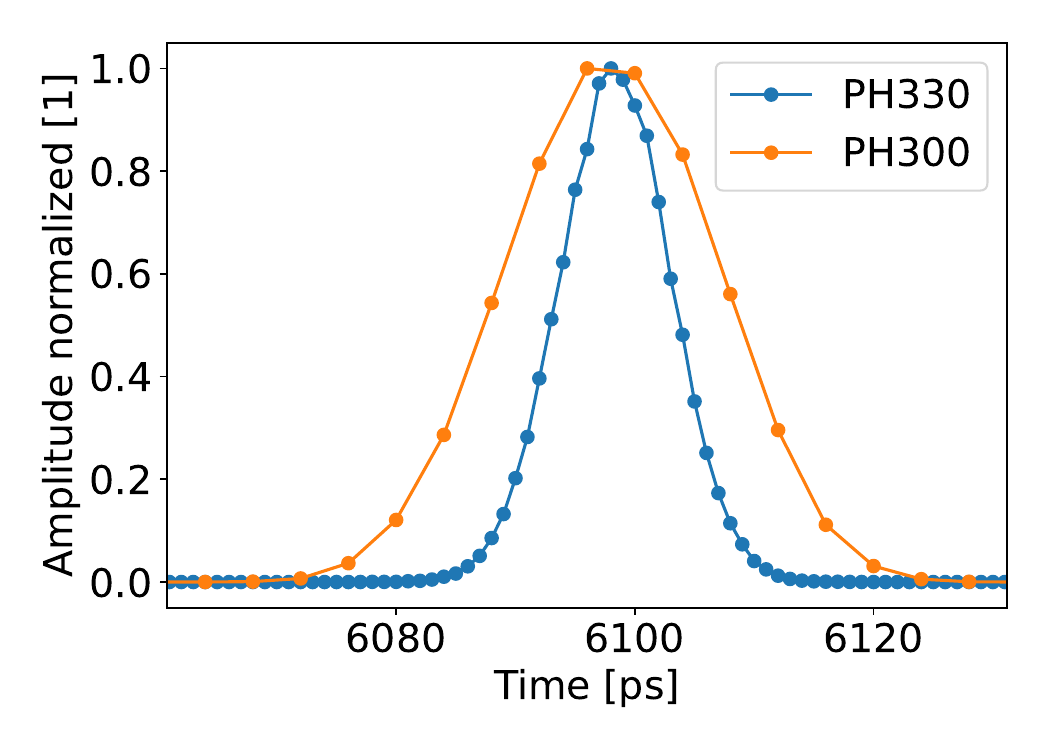}
\vspace{-0.5cm}
\caption{\label{fig:intrinsicjitter} Intrinsic jitter from the PicoHarp~300 and PicoHarp~330 measured by splitting the external trigger to the trigger input and one input channel. The time resolution is the highest possible (PH300: 4\,ps, PH330: 1\,ps).}
\end{wrapfigure}

As histogramming devices a PicoHarp~300~\cite{picoquant_picoharp_2014} as well as a PicoHarp~330~\cite{picoquant_picoharp_2024} are used. One input channel of the histogramming device is connected to the revolution trigger signal which serves as time reference for the measured arrival-time of the counted photon-detection events.
For MAX IV this trigger has a frequency of $\approx$3.125\,MHz (1.5\,GeV ring) and $\approx$0.568\,MHz (3\,GeV ring) which is provided from the laboratory wide timing distribution system via a local event receiver box.
The second channel of the histogramming device is connected to one of the single-photon detectors listed below.
In case of the PicoHarp~330 a third channel is available which can be used to collect data from a second detector in parallel.
The PicoHarp~300 has a maximum of 65536 histogram bins with a minimal temporal bin width of 4\,ps which can be increased in factors of 2.
The internal timing jitter was measured to have a FWHM of 20.7\,ps and was measured by splitting the reference input signal to both input channels. 
The resulting single peak in the histogram can be seen in figure~\ref{fig:intrinsicjitter}.
Both the 1.5\,GeV storage ring as well as the 3\,GeV storage ring have an RF-frequency of $\approx$100\,MHz, corresponding to a bunch spacing of 10\,ns.
Due to the different circumferences the harmonic number (number of stored electron bunches) is 32 for the 96\,m long 1.5\,GeV ring and 176 for the  528\,m long 3\,GeV ring.
To be able to map all stored bunches, a bin width of 8\,ps is used at the 1.5\,GeV ring and a bin width of 32\,ps for the 3\,GeV ring. 

Recently, an upgraded version called PicoHarp~330 has been released.
In the following, some of the new features which can be advantageous for presented measurements are discussed.
One change is that now it is possible to read out two detectors simultaneously. Possible applications for this feature are mentioned in sections~\ref{sec:applications}~and~\ref{sec:profile}.
Furthermore, the new version comes with a higher temporal resolution in the histogram, with the minimal bin width reduced from 4\,ps to 1\,ps.
Additionally, the maximal number of bins was increased to 524288 bins.\footnote{Only available when using the DLL or shared object interface, not available in the official GUI.}
The combination of these two features makes it possible to lower the used resolution to 1\,ps and 4\,ps for the 1.5\,GeV and the 3\,GeV ring, respectively.
The new version also comes with a significantly reduced dead time of less than 1\,ns compared to approximately 85\,ns in the pervious version. This significantly reduces the influence of the histogramming device on dead-time effects in filling-pattern measurement as discussed in, e.g.,~\cite{patting_dead-time_2007, brosi_studies_2017, kehrer_filling_2018}, and shifts the main contribution to dead-time effects towards the single-photon detectors. 
Last but not least, the internal timing jitter in the PicoHarp~330 is reduced.
Measurements resulted in a timing jitter of 10.9\,ps (FWHM) as shown in figure~\ref{fig:intrinsicjitter}.
This directly reduces the system transit-time-spread which is of significance for recovering the longitudinal bunch profiles as is discussed in section~\ref{sec:profile}.

\subsection{Tango device server}

A Tango Controls~\cite{tango_controls} device server was developed to provide a control interface to the histogramming device as well as provide a live analysis of the measured histograms. The software is written in Python and is based on a Cython wrapper and a Tango device developed at ALBA~\cite{blanch-torne_control_nodate}, to interface with the shared object driver provided by the manufacturer. The software was converted to Python3 and extended to also interface with the new PicoHarp~330 and its new features (described above).

\begin{figure}[]
\centering  
\includegraphics[width=0.705\textwidth]{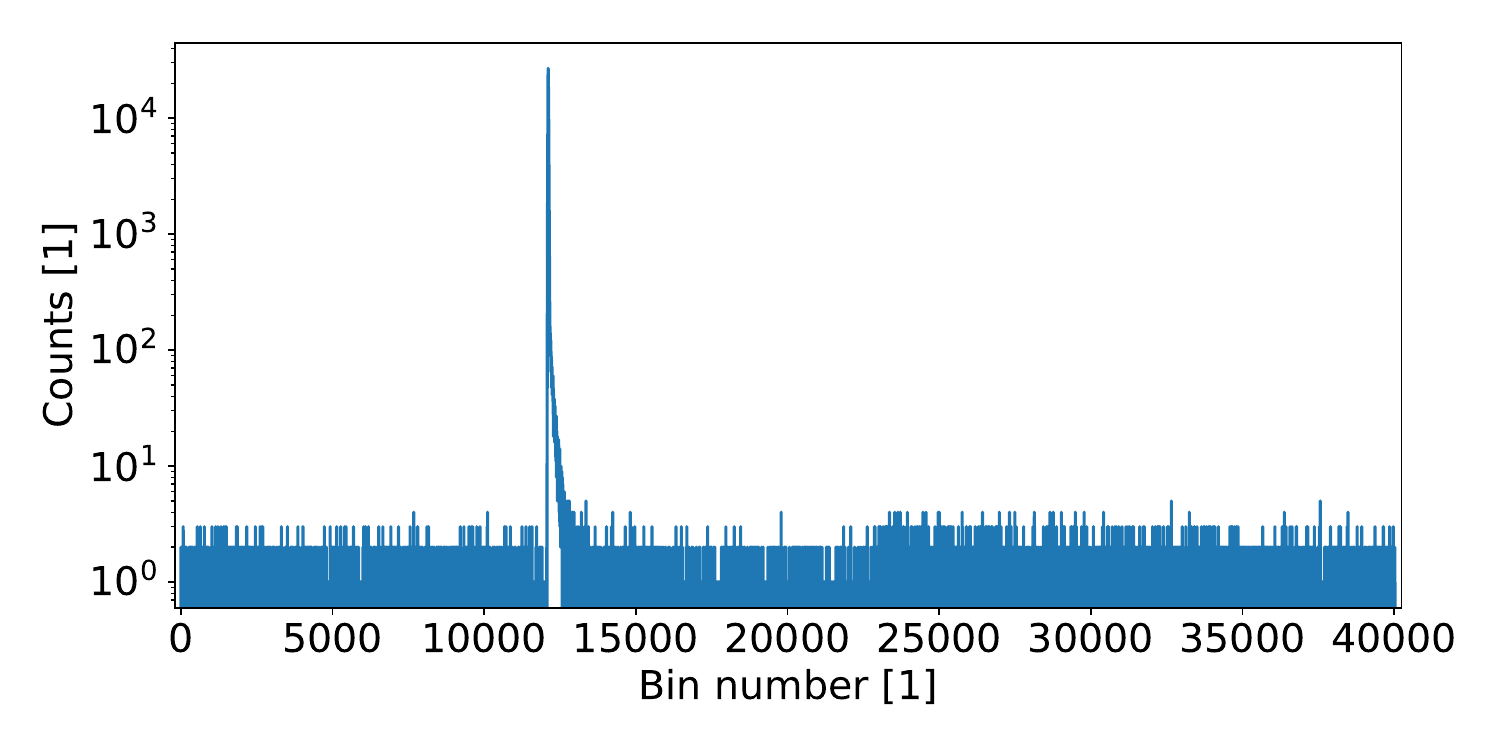}
\includegraphics[width=0.285\textwidth]{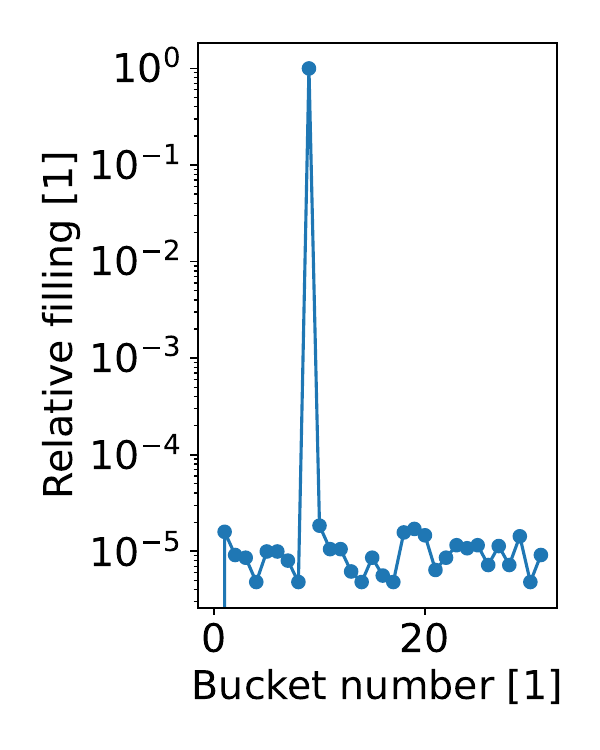}%
\caption{Example measurement of a single bunch in the 1.5\,GeV ring. The histogram on the left shows a purity of approximately $10^4$. With background corrections in the calculation of the filling pattern on the right a purity in the order of $5\cdot10^{4}$ is achieved.  \label{fig:purity}}
\end{figure}

Additionally, an analysis function was added to calculate the filling pattern from the measured histogram directly in the Tango devices. The analysis function allows the correction of background noise by determining a background level based on the ``dark'' areas between the bunches. This increases the measurable purity as shown in figure~\ref{fig:purity}.
The analysis function finds the bunches within the histogram based on the known RF-period of 10\,ns at MAX IV.
After the background correction, the counts for all the bins in each bunch are summed up and normalized by the total number of counts. The resulting array represents the filling pattern which gives the relative portion each bunch holds from the total beam current stored in the ring.

Several derived properties were added to be provided by the Tango device, such as, the homogeneity of the filling pattern when operating with all buckets filled or the purity in case of single-bunch operation.
Furthermore, the updated Tango device server now also calculates the longitudinal bunch profiles and, derived from this, the bunch lengths and phases, as will be described in detail in section~\ref{sec:profile}.

\section{Application of filling pattern measurements at MAX IV\label{sec:applications}}

\begin{figure}[]
    \captionsetup[subfloat]{captionskip=-5pt}
\subfloat[]{
	\includegraphics[width=0.49\textwidth]{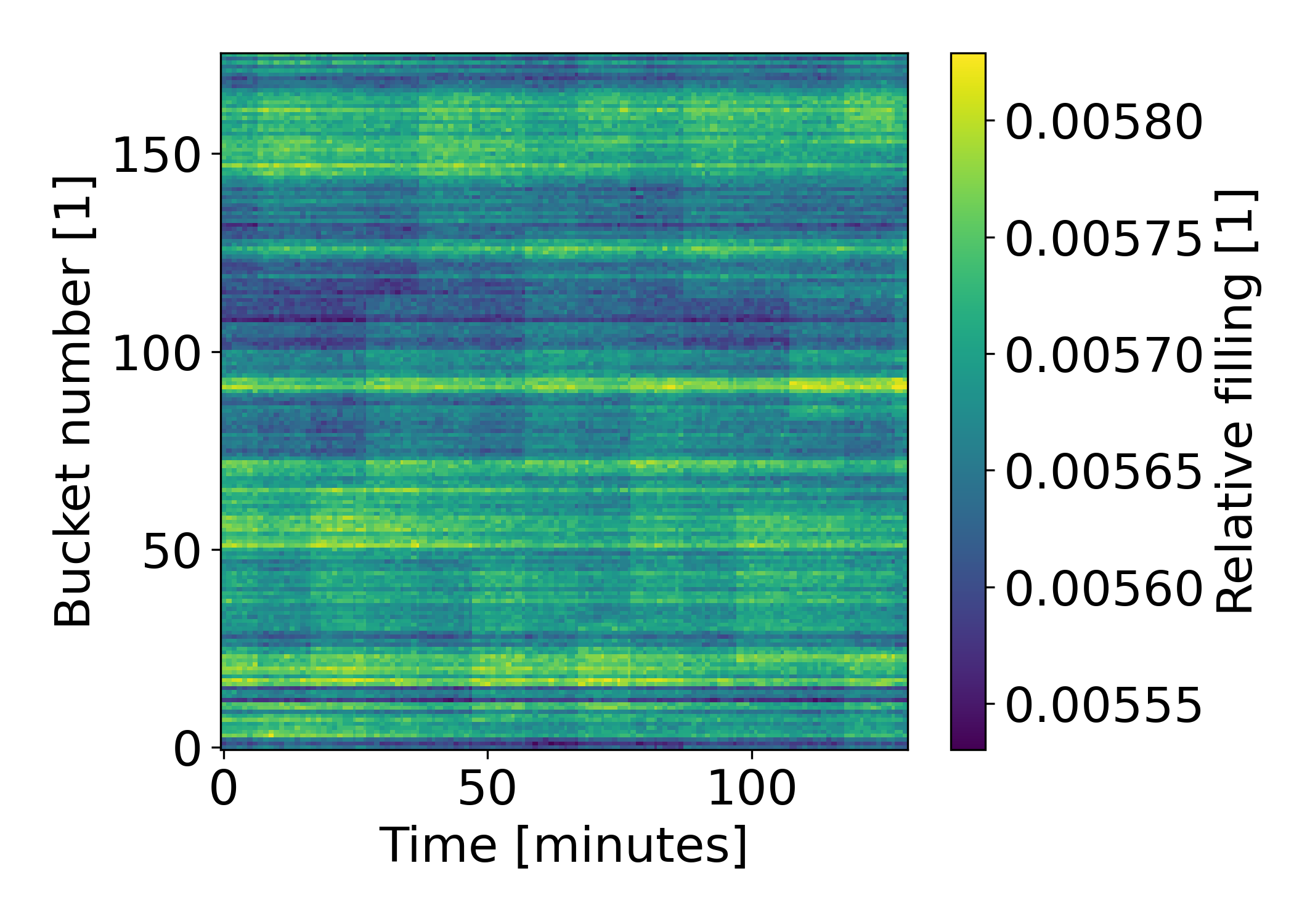}\label{fig:topup}}
\subfloat[]{
	\includegraphics[width=0.49\textwidth]{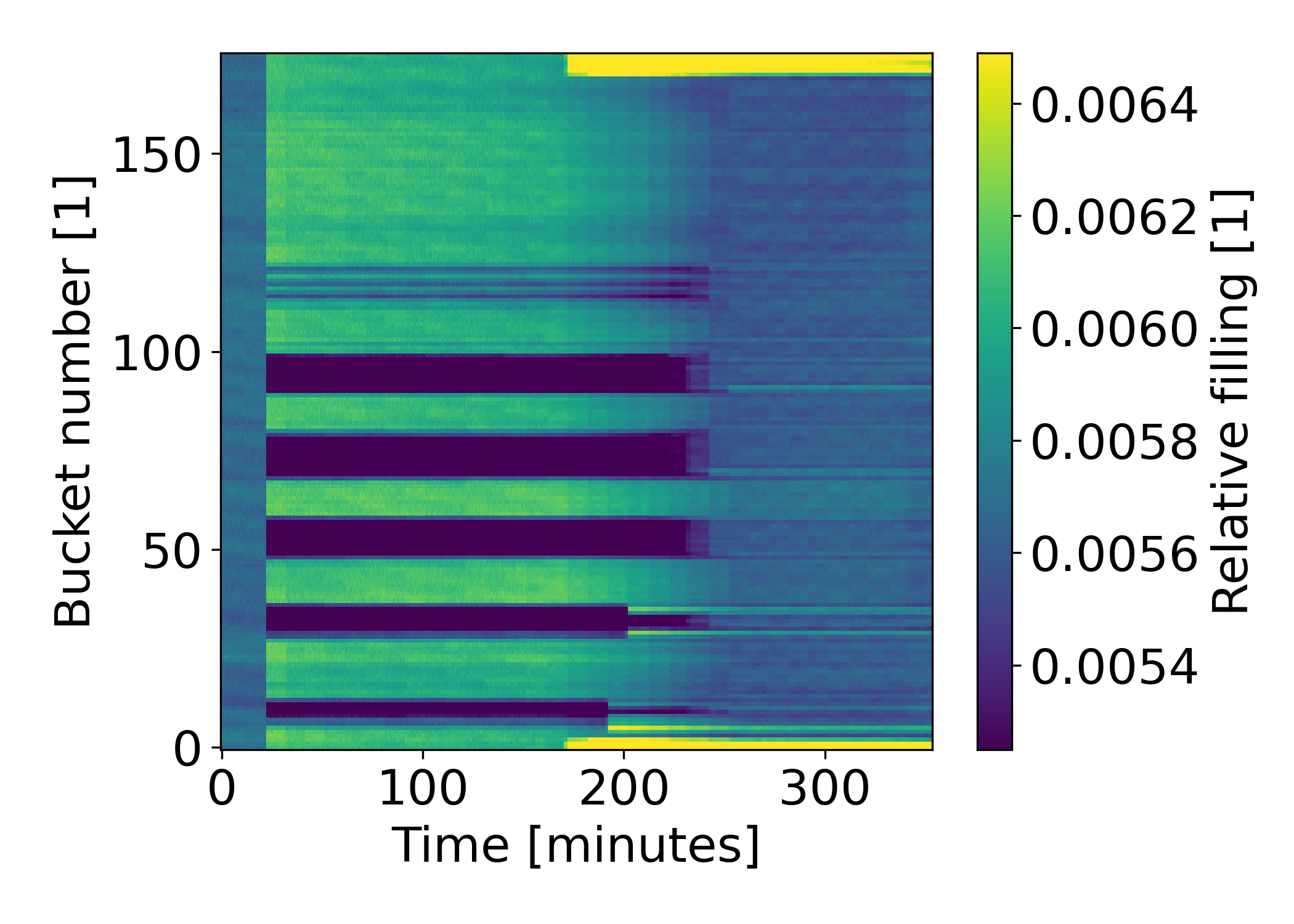}\label{fig:FB}}
%
\caption{Normalized filling pattern as function of time. (a) A slightly inhomogeneous filling pattern that is altered every 10 minutes by the regular top-up injections.  (b) After a partial beam loss (at minute 22), manual efforts were taken to compensate in 4 consecutive top-ups ($\approx$minutes 170--205), leading to overshoots in individual bunches. Filling pattern feedback filled the gaps within 2 top-ups (minutes 230--245).}
\end{figure}

During standard operation at MAX IV, all bunches are filled homogeneously with no gaps in the filling pattern in both rings. 
In this way, all the bunches experience the same lengthening effect by the passive Landau cavities with a resonance frequency at the 3$^\mathrm{rd}$ harmonic of the main RF cavities~\cite{skripka_commissioning_2016}.
A typical value for the homogeneity $H$ during delivery is above 0.990.
It is calculated as unity minus the root-mean-square distance with respect to the average relative filling ($=1 / N_\mathrm{b}$) of the bunches in case of perfect homogeneity. 
In this way, the homogeneity $H$ equals unity for the perfect distribution
$$H=1 - N_\mathrm{b} \sqrt{\sum_{i=0}^{N_\mathrm{b}} \frac{(f_i - (1 / N_\mathrm{b}))^2}{N_\mathrm{b}}},$$
where $N_\mathrm{b}$ is the total number of bunches and $f_i$ is the relative filling in bunch number $i$. 
Due to the continuous losses of a small number electrons, regular top-ups are performed every ten minutes, where a small amount of additional electrons is injected to compensate for the losses. 
The resulting, small changes in the homogeneity (less than 0.5\,\textperthousand~variation) can be resolved with the filing pattern measurement as shown in figure~\ref{fig:topup}.
Even though not every bunch receives additional electrons during one top-up, the homogeneity stays sufficiently high to not require any specific compensation.

In the rare event of a partial loss of the stored electron beam, the standard top-up mechanism might not be sufficient to compensate a potentially resulting inhomogeneity.
Based on the measured filling pattern, it is possible to restore the homogeneity with targeted injections into the bunches affected the most by the previous loss.
An automatic filling-pattern feedback was developed by the operations team. Figure~\ref{fig:FB} shows a demonstration, where the homogeneity was recovered after a partial loss.
In this example, first attempts were conducted manually (within the limited time slots dedicated for top-ups) leading to an overshoot in some bunches.
After the automated filling-pattern feedback is switched on, the homogeneity is recovered within two top-ups (10 minute spacing), with a top-up having a maximal allowed duration of 90 seconds. 
One limitation is that no current is removed by the feedback, so that the manually caused overshoots in some bunches can not be corrected while staying below the allowed total beam current. 
The application of the filling-pattern feedback is faster and less invasive than removing all the remaining beam after a partial loss and filling from scratch to achieve a homogeneous filling pattern again. 

\begin{figure}[b]
\centering
\includegraphics[width=0.485\textwidth]{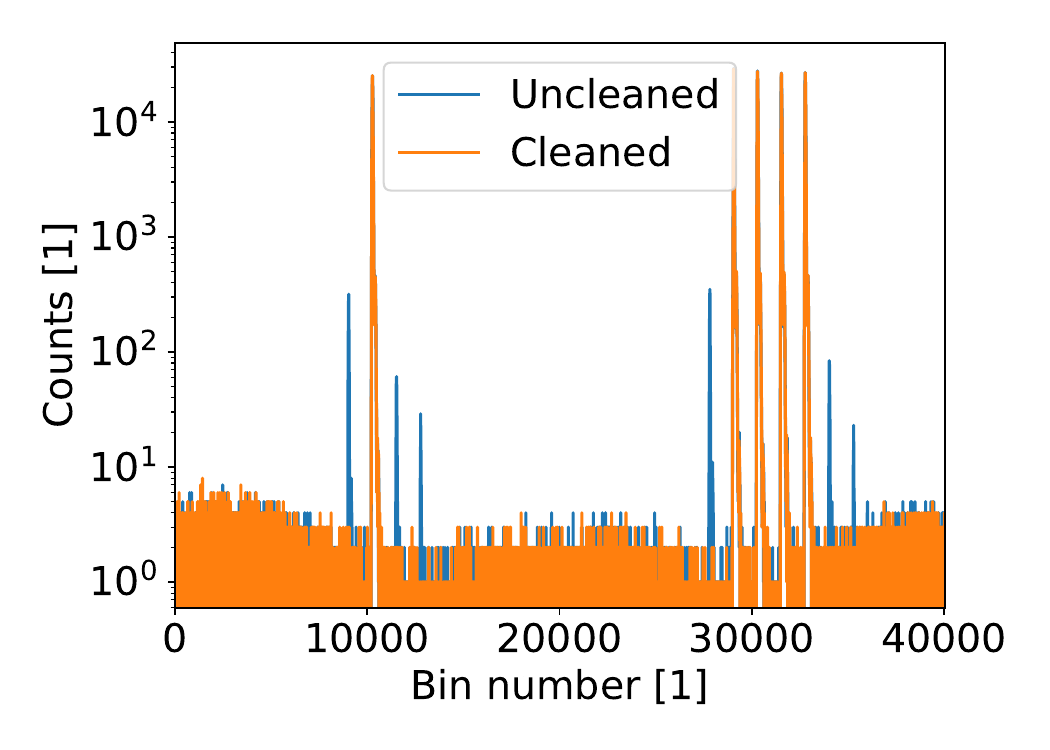}
\includegraphics[width=.485\textwidth]{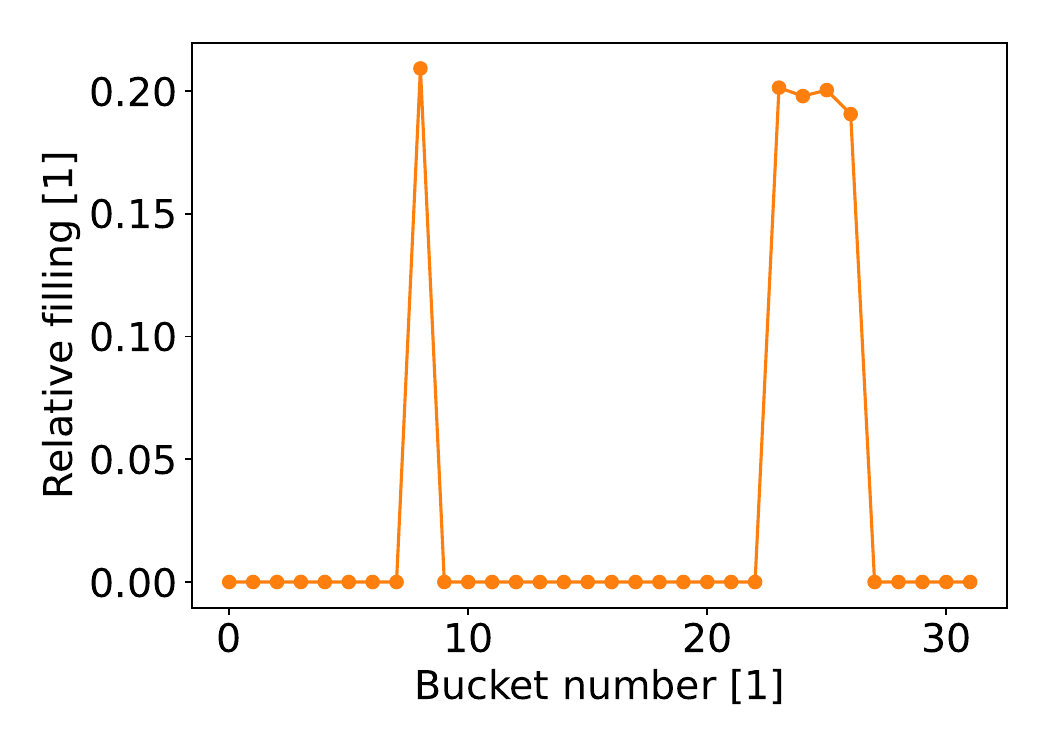}
\caption{Example measurement of a hybrid filling pattern with 5 bunches in the 1.5\,GeV ring before (blue) and after (orange) removing unwanted residual bunches. Left, the measured histograms in logarithmic scale and, right, the calculated filling pattern after cleaning of the unwanted bunches in linear scale. \label{fig:cleaning}}
\end{figure}

Besides the standard operation, a single-bunch mode is provided at the 1.5\,GeV ring as well as custom operation modes which are run for dedicated accelerator physics studies. 
For the single-bunch delivery, a high purity of the filling pattern is important, meaning that no buckets besides one contain electrons. 
The purity can easily be determined from the filling pattern measured via the TCSPC setup. 
The single-bunch purity is then calculated as the ratio between the relative filling in the highest bunch and in the second-highest bunch. 
This gives a value of about unity in multi-bunch delivery and a typical value in the order of $10^4$ in single bunch delivery (for an example see figure~\ref{fig:purity}).
This ratio serves as an important figure of merit for bunch purity during beam delivery in this operation mode, and is also used when optimizing the cleaning procedure that is regularly deployed to remove electrons out of unwanted bunches.

In general, the TCPSC measurement is useful to check the efficiency of bunch cleaning during different operation modes with custom filling patterns, as shown in figure~\ref{fig:cleaning}.
This is, for example, relevant during accelerator physics studies such as measurements of short- or long-range wake fields or synchronous phase shift measurements conducted with and without small witness bunches.
This combined with the mentioned determination of the homogeneity during standard operation and the purity during single-bunch delivery makes the TCSPC measurements an important everyday diagnostic tool.


An additional application of TCSPC measurements was developed at MAX IV. The new method, which allows the determination of the longitudinal bunch profiles, the bunch length and phase of the individual bunches,  will be described in the following section.

\section{Extension to longitudinal bunch profile measurements\label{sec:profile}}

Ideally, a histogram like the one presented in figure \ref{fig:histogramR1} would correspond directly to the filling pattern. There are, however, systematic errors that contribute to the observed profiles of bunches as they appear in the histogram. 
In particular, transit-time differences and timing jitter in relevant components all contribute to distorting the raw histogram bunch profiles.

\subsection{Compensating for the system transit-time spread (TTS)}\label{sec:system_TTS}

In this work, three main limitations to the timing precision of the system were identified,
as indicated in figure \ref{fig:setup}: the detector impulse response function (detector TTS), timing jitter in the revolution clock signal provided by the timing system of the storage ring, and the intrinsic timing jitter of the histogramming device.
To obtain accurate bunch profile measurements from a histogram, the collective impact of these timing jitters needs to be compensated for.
To this end, a method was developed to determine the system TTS, defined here as the total response function of all components involved in TCSPC measurements. 
The following account of the determination of the system TTS and the scheme implemented to compensate for it was developed by the authors in the framework of a thesis project~\cite{schmand_johann_measuring_2024}, which also provides more details for the interested reader.

The bunches as they appear in an untreated histogram are regarded as convolutions of the real bunch profiles with the system TTS. Thus, having determined the system TTS, bunch profiles can be extracted from measured histograms using numerical deconvolution. It should be noted that each different combination of detector, histogramming device, and storage ring is treated as a distinct system, with a corresponding system TTS. Throughout this paper, system TTSs are assumed to be intrinsic properties of a given measurement system, depending only on the hardware used, and not the bunch profiles measured.

The determination of the system TTSs was performed using the TCSPC setup, with the help of a streak camera~\cite{hamamatsu_-__streak_camera_c10910_notitle_nodate}. Assuming a histogram bunch as a convolution of the real bunch profile and the system TTS means that a known bunch shape can be used to determine the system TTS, by deconvolving a histogram bunch with the known bunch shape. A known bunch profile can be achieved by operating the storage ring in low-current mode (in this work, ring currents of 1-3\,mA were used for this purpose). In the low-current limit, bunches approach a Gaussian longitudinal profile (also called \textit{natural} bunch profile). Using the streak camera and simple data processing, the root-mean-square (RMS) bunch lengths $\sigma_{z,0}$ of these natural bunches were determined. For further considerations, an average $\bar\sigma_{z,0}$ over all bunches was used. Using $\bar\sigma_{z,0}$ as its standard deviation, a Gaussian distribution was constructed, acting as an estimated bunch profile for determining the system TTS. The last step to determining the system TTS is to deconvolve histogram bunches using the estimated Gaussian bunch profiles. For the deconvolution, the scikit-image module \cite{scikit-image-deconvolution} with the Richardson-Lucy algorithm \cite{richardson_bayesian-based_1972} in Python was used, an algorithm typically used for image sharpening. It takes two inputs: a signal to be sharpened (in this case a histogram bunch), and a point spread function (here the estimated bunch profile). The Richardson-Lucy algorithm was chosen as the main tool for deconvolution for its numerical stability and noise insensitivity compared to other algorithms. Notably, the Richardson-Lucy algorithm is iterative, and the number of iterations is a parameter to be optimized by the user. First tests indicate that the number of iterations during deconvolution may have a significant impact on the obtained result. For instance, fewer iterations than optimal could yield a broader distribution than a deconvolution using the ideal number of iterations when given the same signal and point spread function. Using the chosen deconvolution algorithm yields one system TTS for each histogram bunch. Once the bunch-specific TTSs were determined, the final system TTS is taken to be the average over all bunches.

It should be noted that the raw histograms are interpolated before deconvolution, increasing the number of data points to the least common multiple of the original number of bins and the ring-specific number of bunches $N_b$. This is done to retain information about relative phases of the bunches. Ensuring that the number of interpolated bins is divisible by $N_b$ allows the interpolated histogram to be divided into equally-sized parts, each part spanning 1/$N_b$ of the revolution time, and each part containing exactly one bunch profile. Interpolation is performed using a cubic spline. 
This interpolation and the subsequent division of the histogram into equally-sized parts avoids systematic offsets between the resulting segments, leading to the desired retention of information about relative phases between bunches.

\begin{figure}[]
    \captionsetup[subfloat]{captionskip=-7pt}
\subfloat[1.5\,GeV ring - PH300 - PMA]{
	\includegraphics[width=0.495\textwidth]{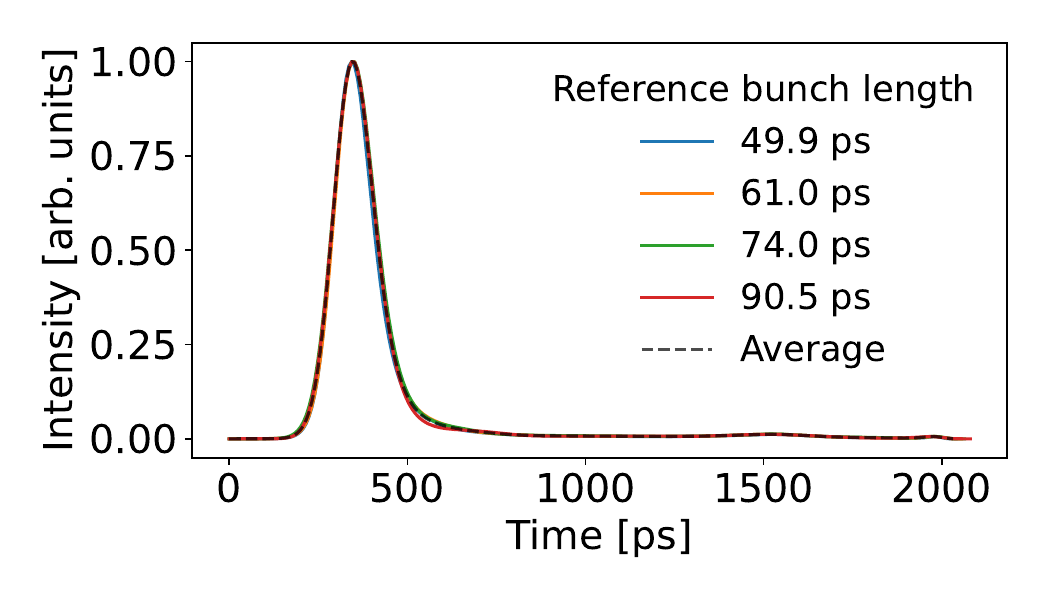}}
\subfloat[1.5\,GeV ring - PH300 - SPAD]{
	\includegraphics[width=0.495\textwidth]{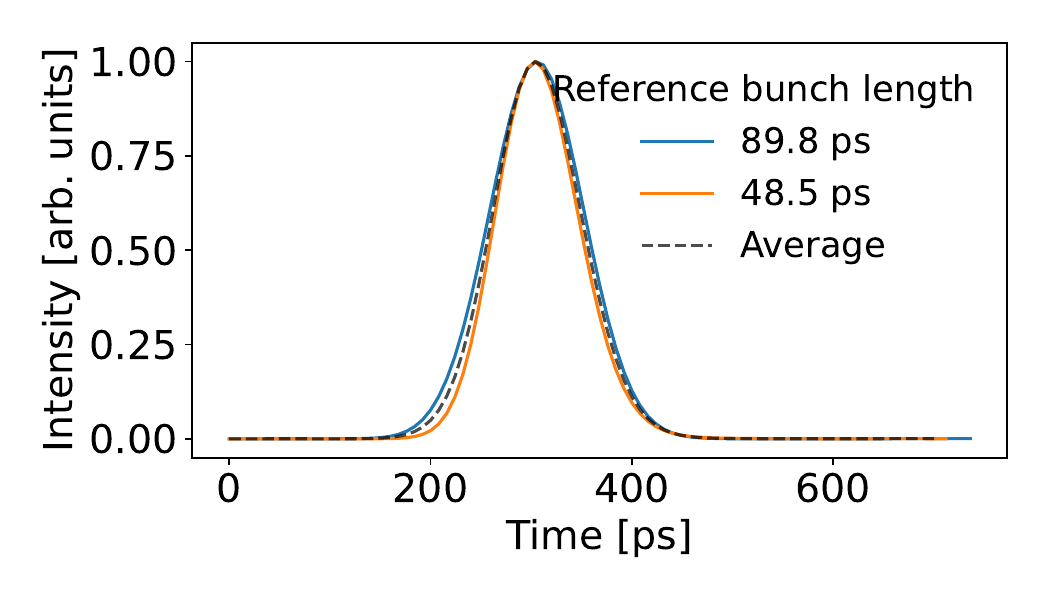}}
\vspace{-0.43cm}
\subfloat[3\,GeV ring - PH300 - PMA]{
	\includegraphics[width=0.495\textwidth]{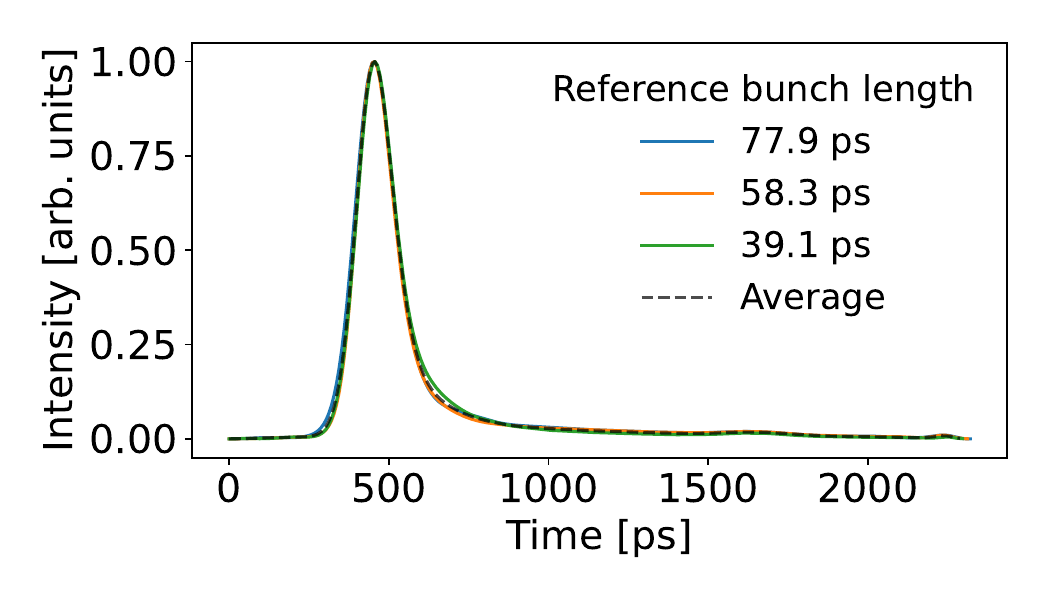}}
\subfloat[3\,GeV ring - PH300 - SPAD]{
	\includegraphics[width=0.495\textwidth]{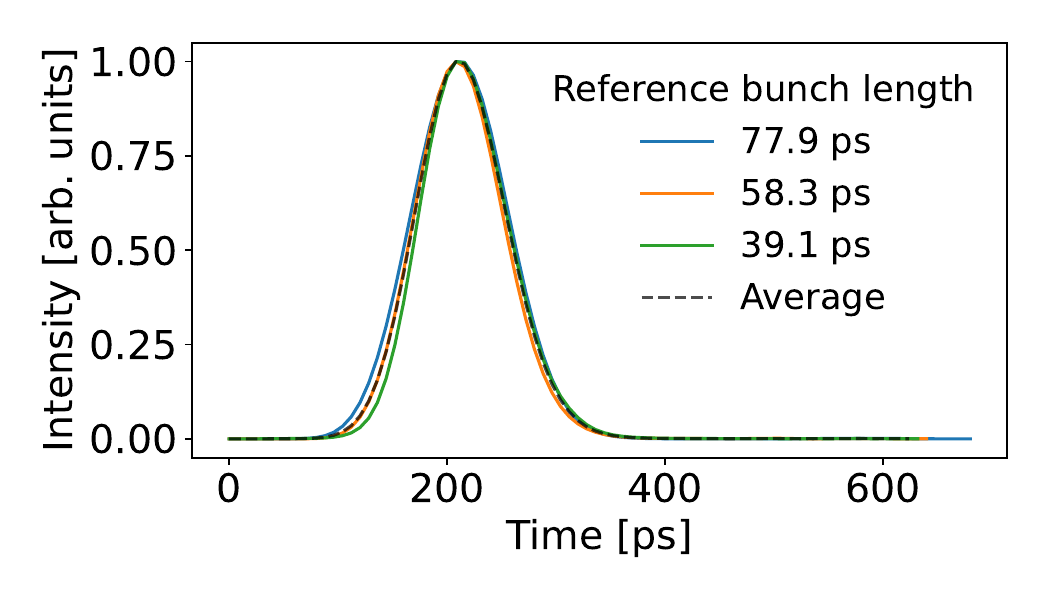}}
\vspace{-0.43cm}
\subfloat[3\,GeV ring - PH330 - PMA Hybrid]{
	\includegraphics[width=0.495\textwidth]{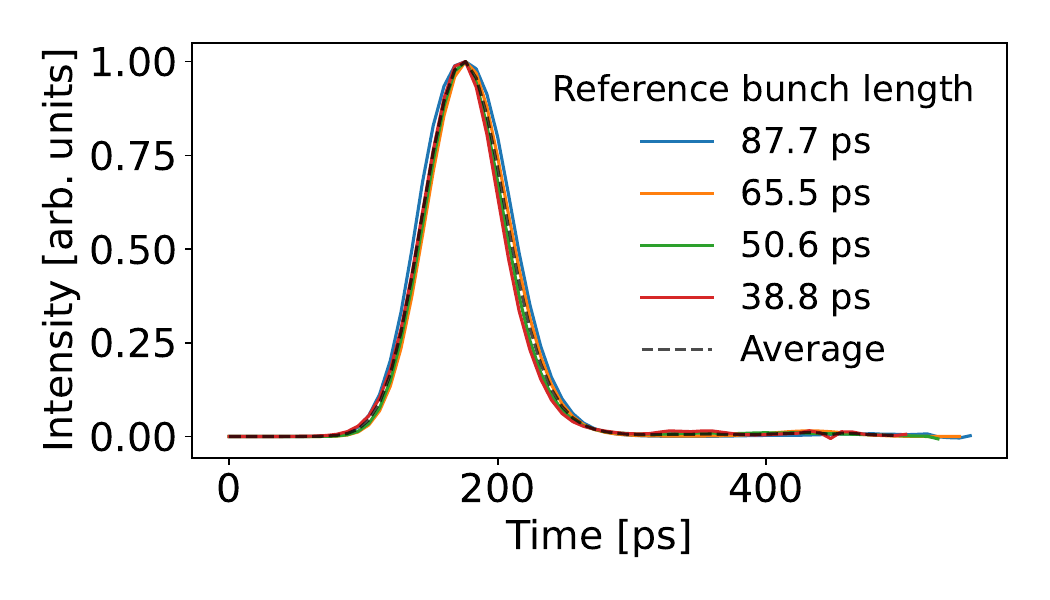}}
\subfloat[3\,GeV ring - PH330 - SPAD]{
	\includegraphics[width=0.495\textwidth]{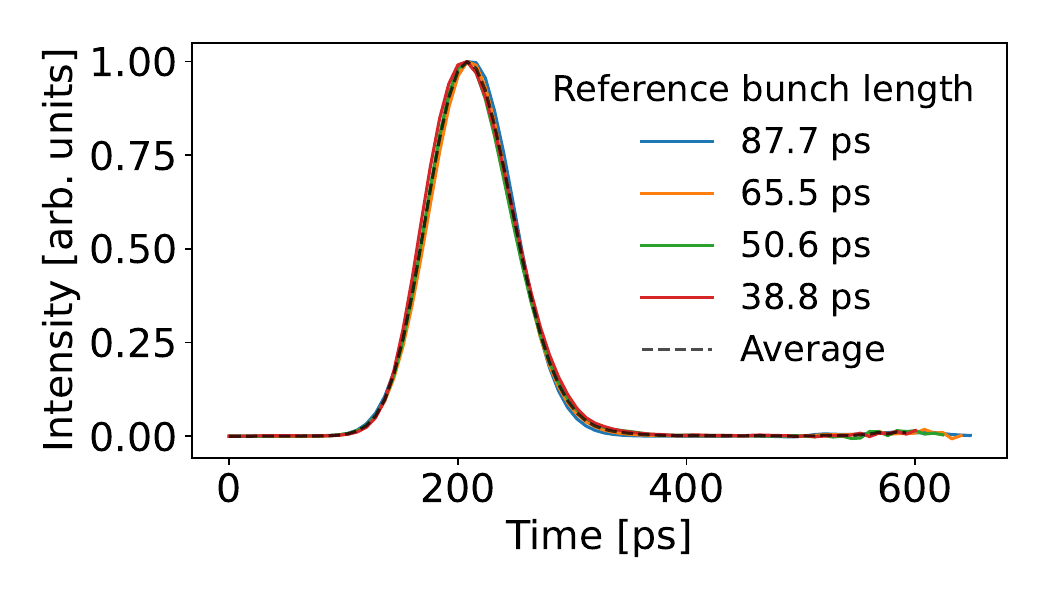}}
\vspace{-0.2cm}
\caption{The determined system TTSs for different combinations of storage ring (1.5\,GeV or 3\,GeV), detector (PMA, SPAD, or PMA Hybrid), and histogramming device (PH300 or PH330). The legend in each plot shows the natural bunch length (rms) used to construct a Gaussian distribution for deconvolving the measured histogram. The time axes have different ranges depending on the system. \label{fig:final_TTSs}}
\end{figure}

Furthermore, the natural bunch length $\sigma_{z,0}$ can be varied by changing the RF-cavity voltage. Thus, determination of system TTSs could be carried out using more than one reference  bunch length. Different runs, each with a different RF-cavity voltage and thus natural bunch length, were carried out at the two storage rings with different detectors. With the assumption that system TTSs are bunch-length-independent, these measurements should yield the same TTS for a given measurement system. Figure \ref{fig:final_TTSs} shows the determined system TTSs for different combinations of storage ring (1.5\,GeV or 3\,GeV), detector (PMA, SPAD, or PMA Hybrid), and histogramming device (PH300 or PH330) determined at different natural bunch lengths. The system TTSs differ slightly depending on the natural bunch length used. 
One possible contributing factor is the accuracy of the bunch length determination via the streak camera. 
To this end, the system TTSs determined at different RF voltages were averaged for each combination of detector, storage ring and histogramming device to obtain a robust solution. 

Once the system TTS has been determined, bunch profiles can be reconstructed from measured histograms via deconvolution, based on the previously determined system TTS, using the same algorithm as described earlier. With the difference, that now the system TTS is deconvolved from the raw histogram to obtain the 
real bunch profile for each bunch individually.

\begin{figure}
\centering
\includegraphics[width=0.49\textwidth]{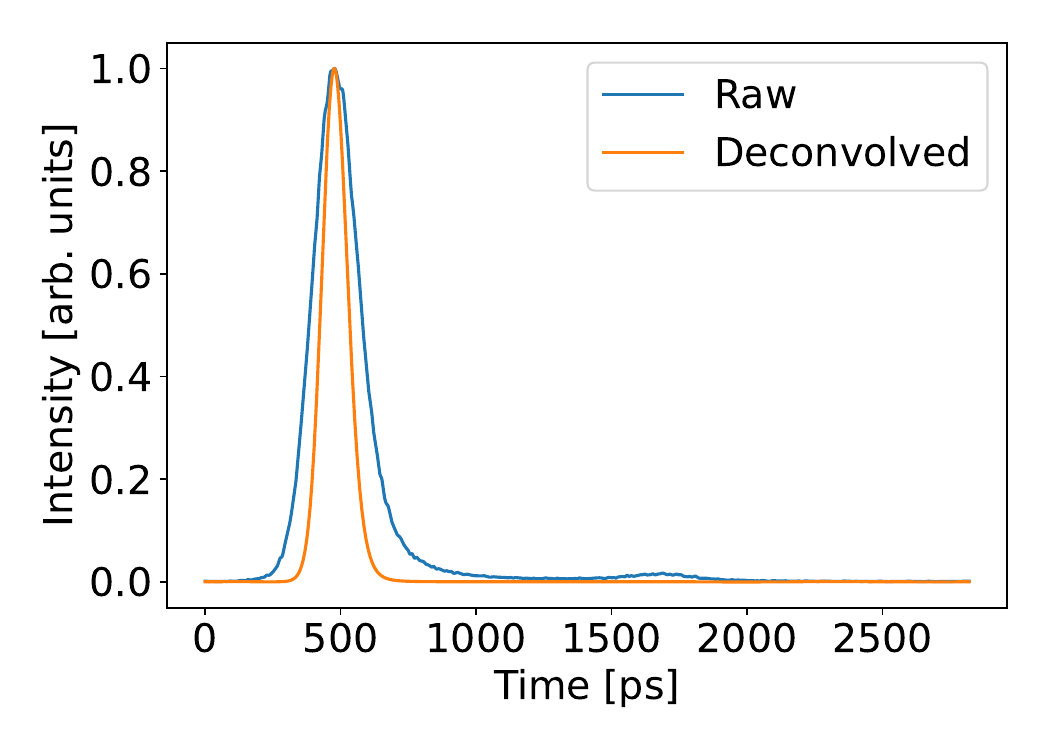}
\includegraphics[width=0.49\textwidth]{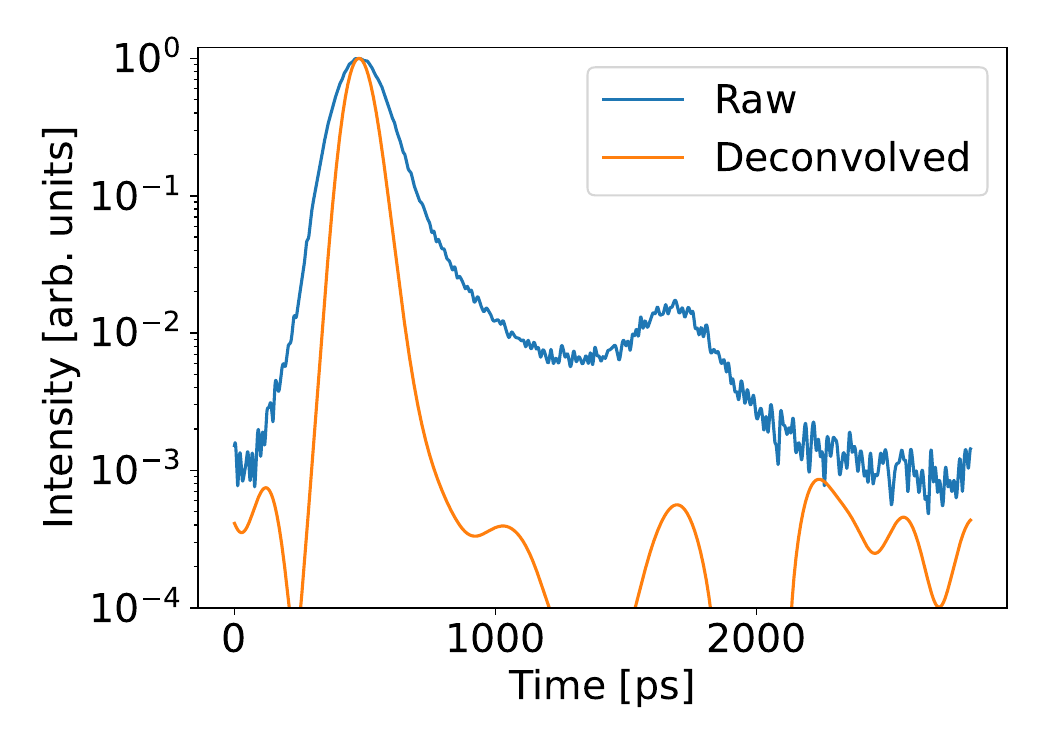}
\vspace{-0.3cm}
\caption{Measured raw histogram distribution for a single low-current bunch at the 1.5\,GeV ring and the resulting distribution after deconvolution. The reduction in length and the resulting Gaussian shape is visible in the linear scale shown on the left side. In the logarithmic scale on the right side the suppression by an additional order of magnitude of the after-pulsing "shoulder", caused by the TTS, is clearly visible.
\label{fig:deconv_single_bunch}}
\end{figure}

An example of such a deconvolution is shown in figure~\ref{fig:deconv_single_bunch}. The measurement with the PMA detector of a low-current bunch in the 1.5\,GeV ring was deconvolved with the system TTS.  
The resulting pulse is shorter as the different contributing timing-jitters and the detector TTS, summarized in the system TTS, are removed. 
Furthermore, it is visible that the deformation of the TTS caused by the after-pulsing of the PMA detector is reduced by more than an order of magnitude, down to approximately the noise level of the shown measurement. 
The resulting pulse is Gaussian as expected for a low-current bunch and the resulting rms pulse length of 50.6\,ps matches the theoretically estimated bunch length of 50.8\,ps for the used machine parameters (in this example an RF voltage of 520\,kV was used). 
The success in recovering the actual bunch profiles via deconvolution can also be seen very well in comparison with streak camera measurements as described in the following section.  

\begin{figure}
    \centering
    \hspace{-0.3cm}\includegraphics[width=0.405\textwidth]{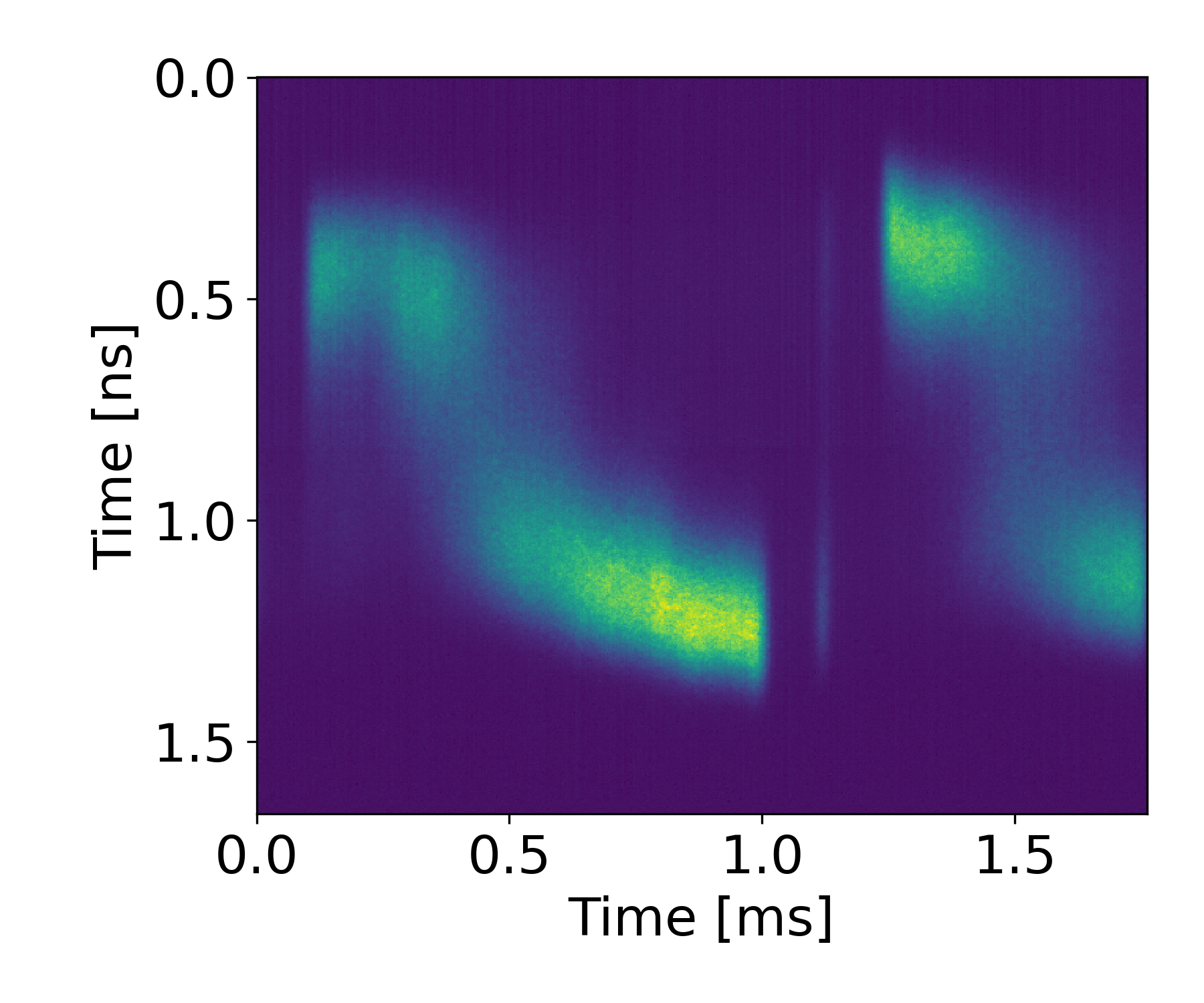}\hspace{-0.3cm}
    \includegraphics[width=0.405\textwidth]{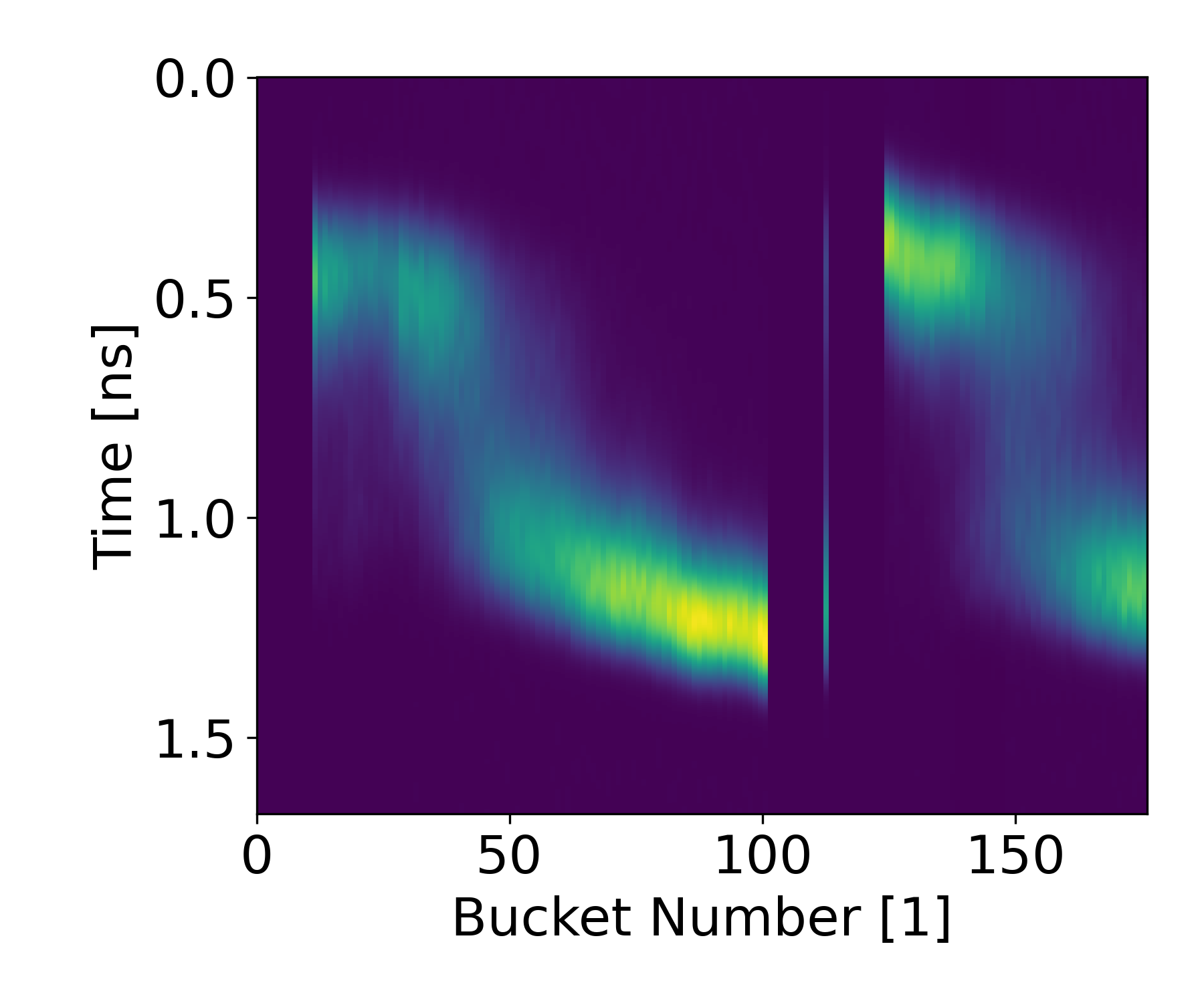}\hspace{-0.3cm}
    \includegraphics[width=0.232\textwidth]{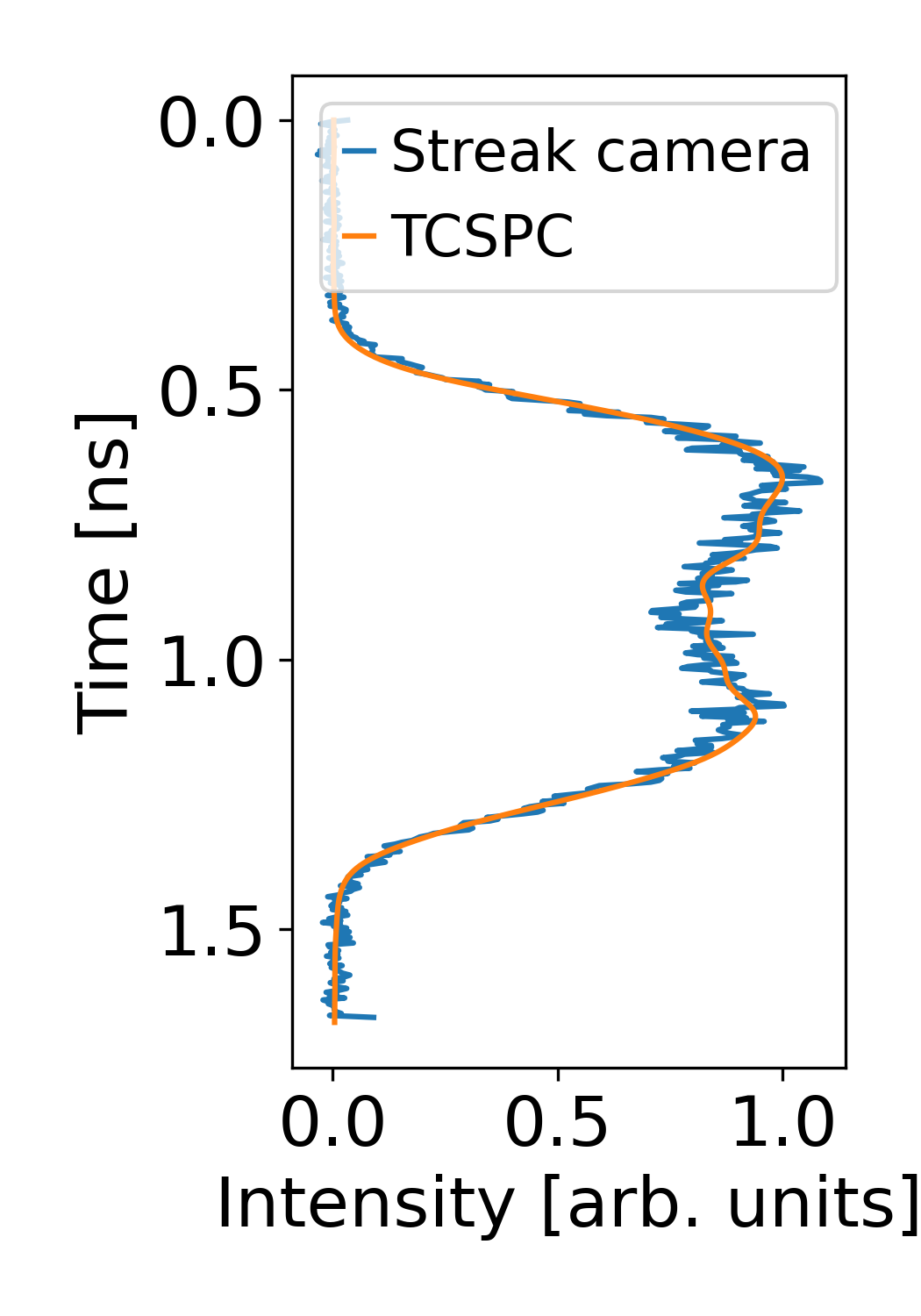}
    \vspace{-0.4cm}
    \caption{Longitudinal bunch profiles measured with an inhomogeneous, high current (400 mA) filling pattern at the 3\,GeV ring, using a streak camera (left) and the TCSPC setup (middle). The TCSPC data has been interpolated and deconvolved using the method described in section \ref{sec:system_TTS}. On the right, the normalised profile of one overstretched bunch (bucket number 44) is displayed as measured with the streak camera and TCSPC.
    }
    \label{fig:comp_sc_TCSPC}
\end{figure}

\subsection{Comparing TCSPC bunch profile and streak camera measurements}\label{sec:comp_TCSPC_SC}
At MAX IV Laboratory, dual-sweep streak cameras from Hamamatsu~\cite{hamamatsu_-__streak_camera_c10910_notitle_nodate} serve as well established tool for longitudinal bunch profiles measurements.
The newly implemented TCSPC setups offer several advantages when measuring longitudinal bunch profiles compared to the streak cameras.
Figure~\ref{fig:comp_sc_TCSPC} shows measurements with an inhomogeneous filling pattern carried out by the streak camera (left) and the TCSPC setup (middle) in parallel.
The two measurements are in good agreement in terms of position of the charge maxima and the bunch lengths.
The shown TCSPC measurement has a higher signal-to-noise ratio (approx. 500), which is an improvement compared to the streak camera (approx. 70), which is also visible as noise on the individual bunch profile shown on the right side of figure~\ref{fig:comp_sc_TCSPC}.
This can largely be attributed to the relatively long integration time used in the TCSPC measurement of 60 seconds while the streak camera measurement was recorded with analog integration overlaying 100 individual exposures of one revolution each. 
Another advantage of TCSPC measurements is that light from adjacent bunches does not overlap as it does in the shown streak camera measurement.
This effect becomes visible in the left plot of figure \ref{fig:comp_sc_TCSPC} when considering the single bunch at index 112. 
In the streak camera plot, the bunch appears smeared out horizontally, whereas the TCSPC setup ensures that light from adjacent bunches does not affect measurements in this way. 
The ability to measure longitudinal bunch profiles \textit{individually} is a key improvement the TCSPC setup offers compared to the streak camera measurement.\footnote{Here we want to clarify that this is only a limitation of the streak camera measurement if all 176 bunches at the 3\,GeV ring should be displayed in one image due with the streak camera sensor. Otherwise it is possible to have a faster horizontal streak and look at individual bunches.}
Last, it should be mentioned that the TCSPC setup is more robust to changes in the intensity of synchrotron radiation compared to the streak camera which has a higher risk to be damaged.
The TCSPC setup is intended for near-continuous use during routine and experimental operation of the storage rings, and the hardware components were chosen with durability in mind. 
This robustness is another advantage over the streak camera, which is rather sensitive and needs to be handled with care and some components like its photo-cathode can deteriorate with time making it less suitable as continuous monitory tool. 
With the TCSPC setup in place, it is now possible to save the streak camera for dedicated experimental beam dynamics studies which require time-resolved measurements, as measurements of this kind on timescales of milliseconds or even single revolutions are the one feature that the TCSPC measurements inherently can not provide. 

\begin{wrapfigure}{R}{0.5\textwidth}
\centering
\includegraphics[width=0.49\textwidth]{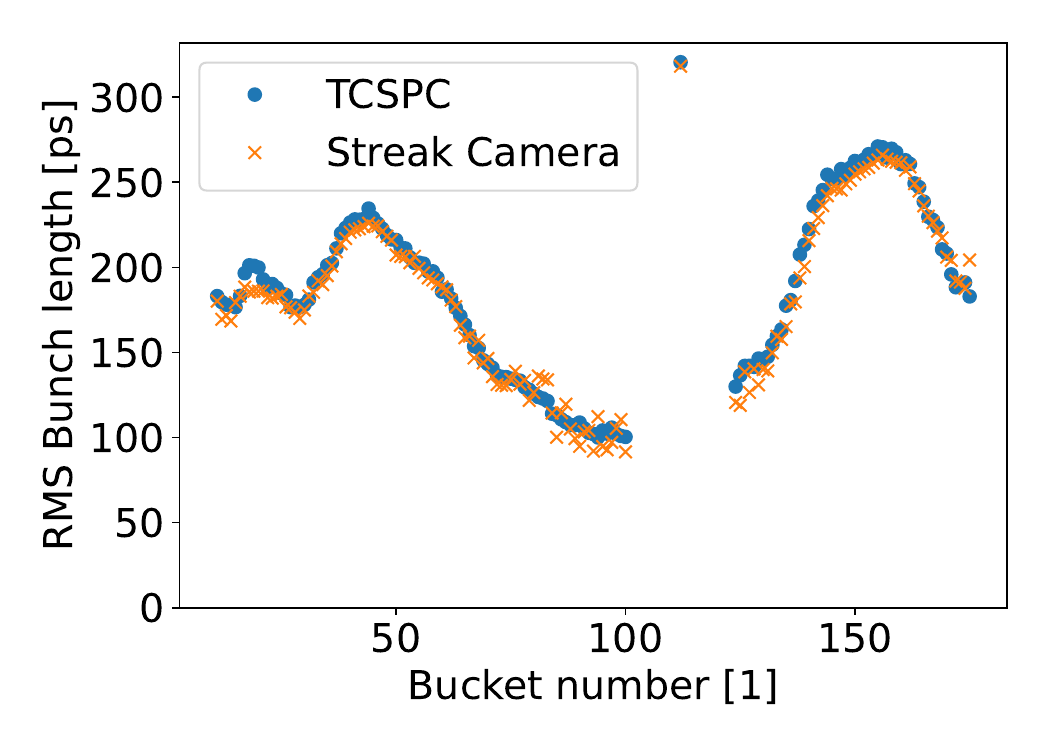}
\vspace{-0.5cm}
\caption{RMS bunch lengths determined from the TCSPC and the streak camera measurements shown in figure~\ref{fig:comp_sc_TCSPC}. The values calculated for empty buckets were removed based on the measured filling pattern. \label{fig:length_landau}}
\end{wrapfigure}

\subsection{TCSPC bunch profile measurements as diagnostic tool}\label{sec:TCSPC_diag}
As mentioned, the TCSPC setup at MAX IV Laboratory is intended as a tool to continuously monitor longitudinal bunch profiles during various modes of operation. 
Two useful parameters which can be obtained from measurements are the bunch length and the synchronous phase, that can be calculated for each bunch separately. 
During normal operation, when all bunches are intended to contain the same charge, these parameters can be used to quickly\footnote{Typical integration times range from 10 seconds to one minute.} spot deviations from the intended beam parameters, as these often result in different than the expected bunch lengths, as well as longitudinal instabilities. 

To this end, the Tango device server was extended to additionally calculate the bunch lengths and synchronous phases of each bunch from its individual bunch profile. The bunch length can be either calculated as an rms value or obtained from a Gaussian fit, with the latter being mostly suitable for measurements at low bunch currents. The phase of each bunch is then calculated as the center-of-mass of the bunch profile. 

Furthermore, the TCSPC measurements also make a good candidate to study the behavior of the beam in atypical conditions, such as inhomogeneous filling patterns. 
The storage rings at the MAX IV Laboratory employ Landau (passive harmonic) cavities, the effects of which are most clearly seen in the case of an inhomogeneous filling pattern.
Figure \ref{fig:comp_sc_TCSPC} shows the effects Landau cavities can have on an the individual bunch profiles in an inhomogeneous filling pattern. 
The most visible features are the emergence of phase transients, a gradual shift of bunch peak positions between the gaps. This is accompanied with differences of up to and more than a factor of two in the bunch lengths, as can be seen in figure~\ref{fig:length_landau}.

\begin{wrapfigure}{R}{0.5\textwidth}
\centering
\includegraphics[width=0.485\textwidth]{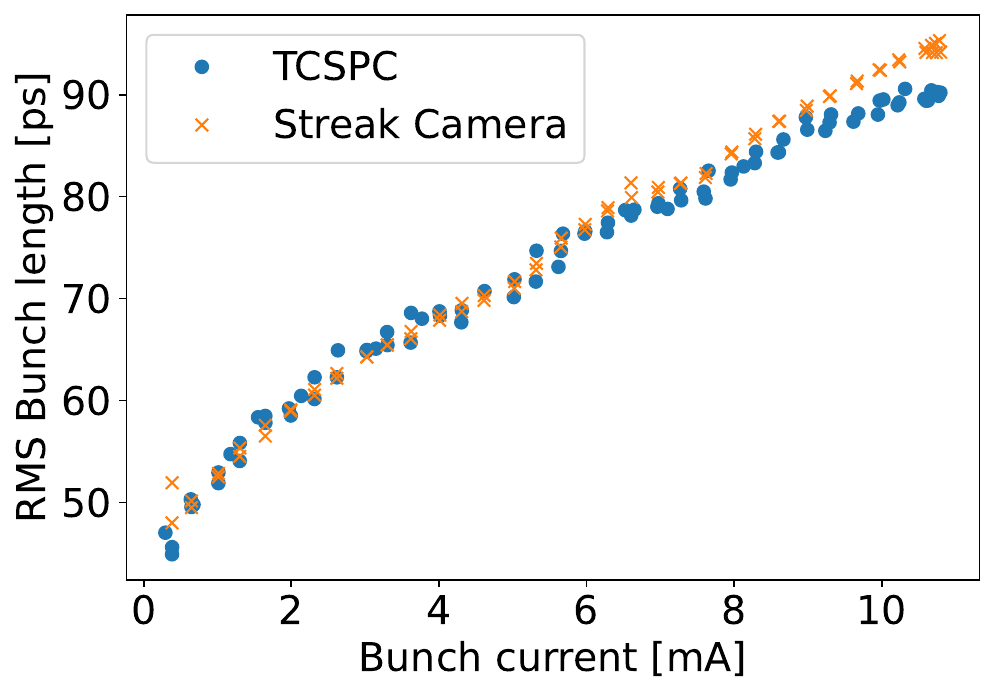}
\vspace{-0.3cm}
\caption{Bunch length, as a function of bunch current, measured with the TCSPC setup and with the streak camera. \label{fig:length_decay}}
\end{wrapfigure}

\begin{figure}
\centering
\captionsetup[subfloat]{captionskip=0pt}
\subfloat[]{
	\includegraphics[width=.485\textwidth]{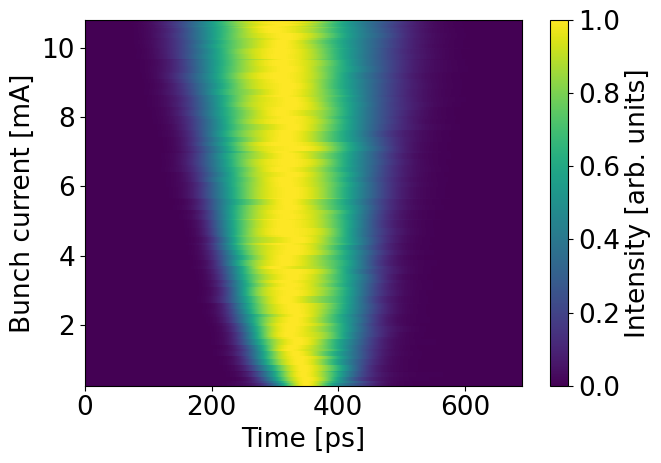}\label{fig:decay2D}}
\subfloat[]{\includegraphics[width=.485\textwidth]{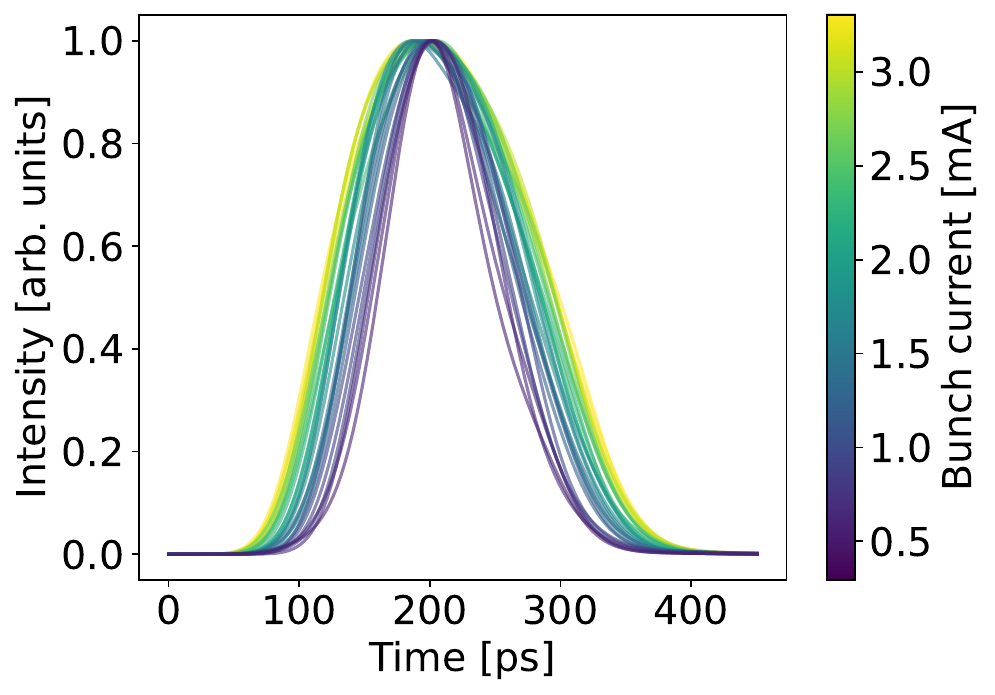}\label{fig:pwd}}
\caption{Bunch profiles measured for a single bunch as a function of bunch current. In (a) the shift in synchronous phase as well as the bunch lengthening is clearly visible. (b) The shown bunch profiles are aligned by their center-of-mass which makes the potential-well distortion visible as the profiles increasingly lean forward (to earlier times) with increasing bunch current.\label{fig:decay}}
\end{figure}

Another example use case for the bunch length and profile measurements are studies of the current-dependent bunch lengthening caused by different impedances such as the resistive-wall impedance or the geometric impedance of the storage ring~\cite{ng_physics_2006}. 
Figure~\ref{fig:length_decay} shows the bunch length as a function of the bunch current. 
There is a good agreement between values determined via TCSPC plus deconvolution and values measured with a streak camera. 
At higher currents, the values diverge slightly, which is attributed to the presence of a micro-wave instability~\cite{ brosi_time-resolved_2023}. 
As the instability results in fluctuations of the longitudinal bunch profile, the determination of a value for the observed bunch length depends on the integration time used in the different measurement types. 
The bunch-lengthening as well as the synchronous-phase shift with current is also very well visible in the bunch profiles shown in figure~\ref{fig:decay2D}.
Furthermore, the expected potential-well distortion caused by the impedances can be seen as deformation of the profiles. 
In figure~\ref{fig:pwd}, the profiles expectedly lean more and more towards the head of the bunch (earlier in time) with increasing bunch current.
For higher bunch currents, the deformation becomes turbulent resulting mainly in a lengthening of the measured profiles.

This example shows again the usefulness of longitudinal bunch profile measurements via TCSPC also for dedicated beam-dynamics studies in addition to standard operation.
As stated in the previous section, the TCSPC setup, due to its robustness and large dynamic range, can be run hands-off which makes it suitable for continuous monitoring.
In our experience, the streak camera, as a highly sensitive and fast diagnostic tool, requires a more careful approach to avoid the risk of over-exposure, especially in conditions with changing bunch current as the experiment shown in figure~\ref{fig:length_decay}.

     \subsection{Limitations}
There are several limitations to the presented method of determining the longitudinal bunch profiles via TCSPC measurements.

Firstly, the chosen deconvolution algorithm works with an iterative approach and the number of iterations is to be given as external parameter. 
For the presented application, the optimal number of iterations was determined by systematic tests as a too low number resulted in an incomplete suppression of the TTS effects and a too high number caused strong overshoots at the edge of the resulting profiles.
The tests showed, that the same number of iterations could be used for all setup combinations of detectors and histogramming devices at MAX IV.
Nevertheless, this should be kept in mind as a possible source of errors on the determined bunch profiles. 

A further limitation is the comparably long integration time necessary to collect a high number of counts in the histogram and by this reduce the error on the measured distribution. 
As a counting experiment, the TCSPC measurement follows Poisson statistics with an relative uncertainty on the number of counts $N$  of $\sigma_N=\frac{\sqrt{N}}{N}$. 
For the resulting filling pattern this gives an uncertainty on the relative filling per bunch of approximately 0.3\% for a 60\,s measurement with a total count rate of about 1 count per revolution ($\approx$567\,kHz) for all 176 bunches in the 3\,GeV ring. 
The uncertainty on the determined bunch profile is higher and can be roughly estimated as $\sigma_N$ times the number of histogram bins covered by a bunch. 
At the same time, in most cases it can be assumed that the bunch profile is rather smooth and does not change significantly between histogram bins. 
This relaxes the demand on the required number of counts somewhat. 
However, it is still important to test for a sufficient number of counts as excessive noise on the measured bunch profile can have unwanted effects in the deconvolution algorithm.
The need for the long integration time is caused by the maximal allowed count rate. 
The main limitation on the count rate is the requirement to measure in the single-photon counting regime, where the probability to detect two photons in one event is negligible.
This ensures the linear correlation between the counting events and the number of electrons in the bunches, which is the fundament for using the TCSPC method for this diagnostic.
The time scales that ensure this are dominated by the used equipment and depend  strongly on the dead-time effect, where the detector or the histogramming device is ``blind'' for a certain time span after an event resulting in missed consecutive events leading to deformations of the measured histogram.
An algorithm to estimate and compensate dead-time effects~\cite{patting_dead-time_2007} has been successfully tested in the past~\cite{kehrer_filling_2018}.

\begin{wrapfigure}{R}{0.5\textwidth}
\centering
\includegraphics[width=0.49\textwidth]{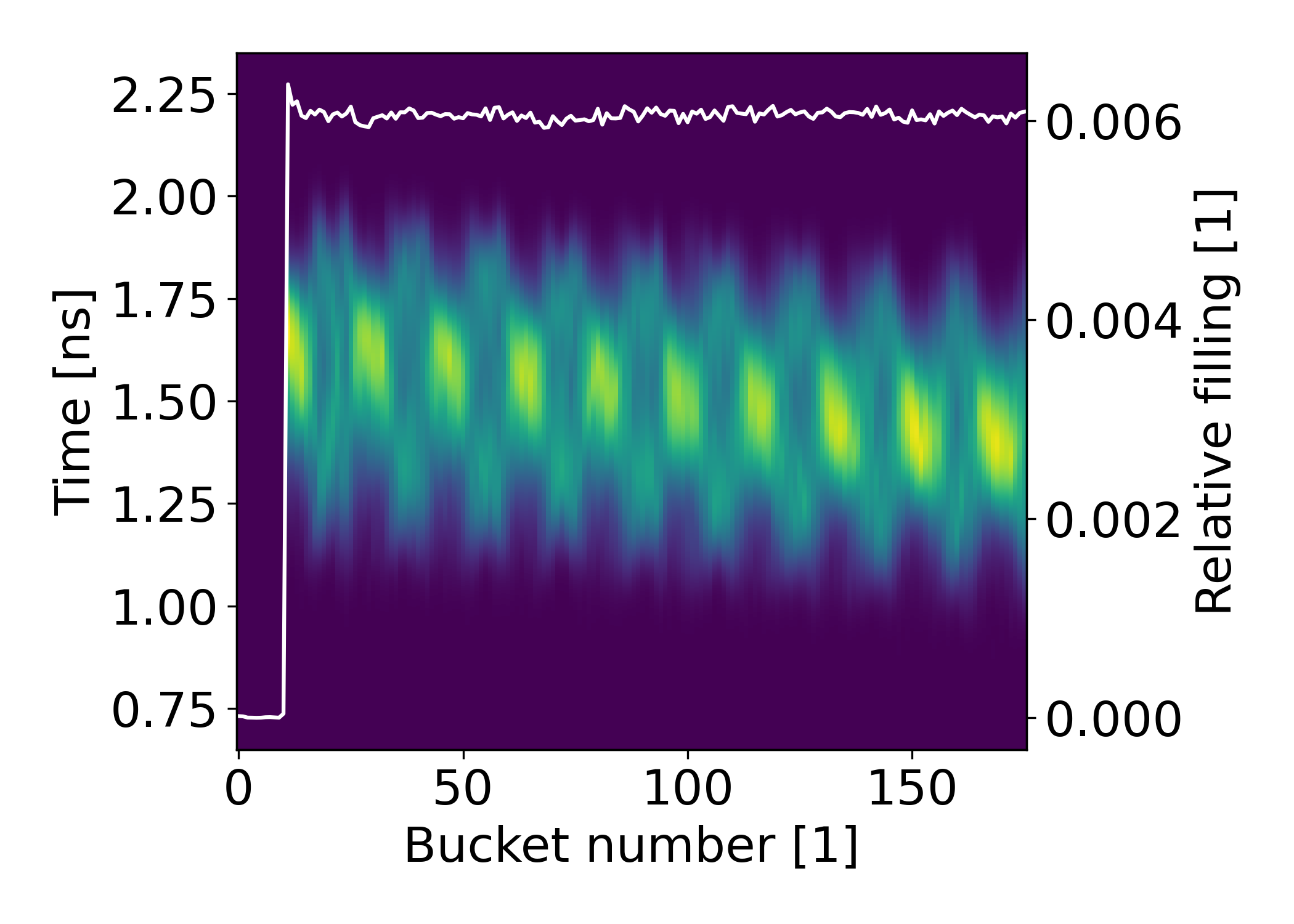}
\vspace{-0.5cm}
\caption{Coupled-bunch modes, as seen in the TCSPC measurement in a homogeneous filling pattern with a 11 bucket gap (white line). The apparent difference in bunch length is mainly caused by the difference in motion of the individual bunches which is averaged due to the long acquisition time. \label{fig:CBM}}
\end{wrapfigure}
The main impact of the long integration time is the inability to resolve dynamics of the electron beam faster than a few seconds.
All faster longitudinal movements and oscillations will, without additional information, be indistinguishable from an increase in bunch length.
An example for this are the bunch movements during coupled-bunch instabilities (e.g.~\cite{chao_physics_1993, ng_physics_2006}).  Figure~\ref{fig:CBM} shows the measured bunch profiles of all bunches in the presence of coupled-bunch modes. 
The filling pattern extracted from the measurement shows a homogeneous fill with one gap of 11 bunches. 
It is visible how the bunches at the nodes of the couple-bunch modes appear shorter as they are moving less than the bunches in between the nodes. 
Albeit, with the additional information from, for example, a bunch-by-bunch beam position monitor indicating the presence of the instability, the measured bunch profiles can be interpreted accordingly and can provide more insight into the dynamics of the instability.

\section{Summary}

The time-correlated singe-photon counting (TCSPC) method is used as standard measurement of the distribution of the charge onto the electron bunches at many storage-ring-based synchrotron light sources. 
Two new setups were installed for the two storage rings at the MAX IV Laboratory and the impact of different detector types and a new version of the histogramming device was discussed.

In addition to the now available continuous monitoring of the filling pattern, the TCSPC setups help with ensuring a high purity in special filling patterns, e.g., during dedicated timing mode operations as well as with maintaining the homogeneity during the operation with harmonic cavities.

Furthermore, the measurement analysis was developed further to allow the determination of the longitudinal bunch profiles of the individual bunches. 
To this end, the transit-time spread of the TCSPC system, including the different timing jitter contributions from the employed detector, the timing distribution system, and the used histogramming device, was determined. 
Via deconvolution, this contribution can be removed from the measured histograms returning the real charge distribution even within the bunches. 
It could be shown that the longitudinal bunch profiles obtained with the developed method agree very well with comparable measurements taken with a streak camera. 
The TCSPC system with its robustness and wide dynamic range with respect to light intensity is a valuable complement to a streak camera, in the role of a continuous monitoring tool for longitudinal beam parameters.

The newly developed method to determine longitudinal bunch profiles proved to be a very useful diagnostic in complex beam dynamics studies and special operation modes, and especially in the operation of a 4th generation light source relying on bunch lengthening via harmonic cavities.

\acknowledgments
The authors would like to thank the KITS team, in particular, Áureo Freitas for the support with the integration of the analysis algorithms into the accelerator control system. 
We also would like to thank Andreas Johansson from the operations group for integrating the TCSPC measurement of the filling pattern into the filling pattern feedback. 
Furthermore, we would like to thank Francis Cullinan from the accelerator development group for his patience and his valuable feedback as beta tester of the bunch profile measurements during accelerator physics studies.
And last, but not least, we thank Åke Andersson for his support.




\bibliographystyle{JHEP}
\bibliography{TCSPC_paper.bib}

\providecommand{\href}[2]{#2}\begingroup\raggedright\begin{thebibliography}{10}

\bibitem{yang_bunch-by-bunch_2010}
B.~Yang, W.~Norum, S.~Shoaf and J.~Stevens, \emph{{{BUNCH-BY-BUNCH DIAGNOSTICS
  AT THE APS USING TIME- CORRELATED SINGLE-PHOTON COUNTING TECHNIQUES}}},  in
  \emph{{{BIW10}}}, (Santa Fe, New Mexico, US), Feb., 2010,
  \href{https://accelconf.web.cern.ch/BIW2010/papers/tupsm044.pdf}{https://accelconf.web.cern.ch/BIW2010/papers/tupsm044.pdf}.

\bibitem{schmand_johann_measuring_2024}
{Schmand, Johann}, \emph{Measuring Longitudinal Bunch Profiles Using
  Time-Correlated Single-Photon Counting at {{MAX IV}} Laboratory}, bachelor
  thesis, Lund University, Lund, Sweden, June, 2024.
\newblock
  \href{http://lup.lub.lu.se/student-papers/record/9164187}{http://lup.lub.lu.se/student-papers/record/9164187}.

\bibitem{wahl_time-correlated_2014}
M.~Wahl, \emph{Time-{{Correlated Single Photon Counting}}},  Tech. Rep.
  \href{https://www.picoquant.com/images/uploads/page/files/7253/technote\_tcspc.pdf}{https://www.picoquant.com/images/uploads/page/files/7253/technote\_tcspc.pdf},
  PicoQant GmbH (2014).

\bibitem{torino_filling_2014}
L.~Torino and U.~Iriso, \emph{Filling {{Pattern Measurements}} at {{ALBA}}
  using {{Time Correlated Single Photon Counting}}},
  \href{https://doi.org/10.18429/JACOW-IPAC2014-THPME162}{\emph{Proceedings of
  the 5th Int. Particle Accelerator Conf.} {\bfseries IPAC2014} (2014) 3 pages,
  0.338 MB}.

\bibitem{torino_time_2017}
L.~Torino and U.~Iriso, \emph{Time {{Correlated Single Photon Counting Using
  Different Photon Detectors}}},
  \href{https://doi.org/10.18429/JACOW-IBIC2016-MOPG59}{\emph{Proceedings of
  the 5th Int. Beam Instrumentation Conf.} {\bfseries IBIC2016} (2017) 4 pages,
  1.450 MB}.

\bibitem{thomas_bunch_2006}
C.~Thomas, G.~Rehm, H.~Owen, N.~Wyles, S.~Botchway, V.~Schlott et~al.,
  \emph{Bunch purity measurement for {{Diamond}}},
  \href{https://doi.org/10.1016/j.nima.2006.07.059}{\emph{Nuclear Instruments
  and Methods in Physics Research Section A: Accelerators, Spectrometers,
  Detectors and Associated Equipment} {\bfseries 566} (2006) 762}.

\bibitem{thomas_time_2007}
C.A.~Thomas and G.~Rehm, \emph{{{TIME DOMAIN MEASUREMENTS AT DIAMOND}}},  in
  \emph{{{DIPAC}} 2007}, (Venice, Italy), JACoW, May, 2007,
  \href{https://accelconf.web.cern.ch/d07/papers/wepb25.pdf}{https://accelconf.web.cern.ch/d07/papers/wepb25.pdf}.

\bibitem{rehm_different_2014}
G.~Rehm, \emph{Different {{Detectors}} for {{Time Correlated Single Photon
  Counting}}},  in \emph{{{DEELS}} 2014}, (Grenoble, France), Dec., 2014,
  \href{https://www.esrf.fr/files/live/sites/www/files/events/conferences/2014/DEELS
  2014/DEELS\_GR\_Detectors\_for
  TCSPC.pdf}{https://www.esrf.fr/files/live/sites/www/files/events/conferences/2014/DEELS
  2014/DEELS\_GR\_Detectors\_for TCSPC.pdf}.

\bibitem{holldack_bunch_2007}
K.~Holldack, M.~V.~Hartrott, F.~Hoeft, O.~Neitzke, E.~Bauch and M.~Wahl,
  \emph{Bunch fill pattern at {{BESSY}} monitored by time-correlated single
  photon counting},  in \emph{Optics {{East}} 2007}, W.~Becker, ed., (Boston,
  MA), p.~677118, Sept., 2007, \href{https://doi.org/10.1117/12.734226}{DOI}.

\bibitem{wu_filling_2008}
C.~Wu, K.~Hu, J.~Chen, C.~Kuo and K.~Hsu, \emph{{{FILLING PATTERN MEASUREMENT
  FOR THE TAIWAN LIGHT SOURCE}}},  in \emph{{{EPAC08}}}, (Genoa, Italy), June,
  2008,
  \href{https://epaper.kek.jp/e08/papers/tupc038.pdf}{https://epaper.kek.jp/e08/papers/tupc038.pdf}.

\bibitem{corbett_bunch_2014}
J.~Corbett, P.~Leong and L.~Zavala, \emph{{{BUNCH PATTERN MEASUREMENT VIA
  SINGLE PHOTON COUNTING AT SPEAR3}}},  in \emph{{{IBIC2014}}}, Sept., 2014,
  \href{https://accelconf.web.cern.ch/ibic2014/papers/mopd21.pdf}{https://accelconf.web.cern.ch/ibic2014/papers/mopd21.pdf}.

\bibitem{cope_upgrades_2016}
T.~Cope, S.~Allison, J.~Corbett and Y.~Xu, \emph{Upgrades to the {{SPEAR3
  Single-Photon Bunch Measurement System}}},  in \emph{Proceedings of the 7th
  {{Int}}. {{Particle Accelerator Conf}}.}, vol.~IPAC2016, pp.~2 pages, 0.711
  MB, JACoW, Geneva, Switzerland, 2016,
  \href{https://doi.org/10.18429/JACOW-IPAC2016-THPOY051}{DOI}.

\bibitem{xu_electron_2018}
B.~Xu, E.~Carranza, A.~Chen, S.~Condamoor, J.~Corbett and A.~Fisher,
  \emph{Electron {{Bunch Pattern Monitoring}} via {{Single Photon Counting}} at
  {{SPEAR3}}},
  \href{https://doi.org/10.18429/JACOW-IBIC2017-TUPCC09}{\emph{Proceedings of
  the 6th Int. Beam Instrumentation Conf.} {\bfseries IBIC2017} (2018) 4 pages,
  0.909 MB}.

\bibitem{kehrer_visible_2015}
B.~Kehrer, A.~Borysenko, E.~Hertle, N.~Hiller, M.~Holz, A.-S.~M{\"u}ller
  et~al., \emph{Visible {{Light Diagnostics}} at the {{ANKA Storage Ring}}},
  \href{https://doi.org/10.18429/JACOW-IPAC2015-MOPHA037}{\emph{Proceedings of
  the 6th Int. Particle Accelerator Conf.} {\bfseries IPAC2015} (2015) 3 pages,
  1.022 MB}.

\bibitem{kehrer_filling_2018}
B.~Kehrer, E.~Blomley, M.~Brosi, E.~Br{\"u}ndermann, A.-S.~M{\"u}ller, M.~Schuh
  et~al., \emph{Filling {{Pattern Measurements Using Dead-Time Corrected Single
  Photon Counting}}},  in \emph{Proc. 9th {{International Particle Accelerator
  Conference}} ({{IPAC}}'18), {{Vancouver}}, {{BC}}, {{Canada}}, {{April}}
  29-{{May}} 4, 2018}, no.~9, (Geneva, Switzerland), pp.~2219--2222, JACoW
  Publishing, June, 2018,
  \href{https://doi.org/10.18429/JACoW-IPAC2018-WEPAL027}{DOI}.

\bibitem{breunlin_emittance_2016}
J.~Breunlin and {\AA}.~Andersson, \emph{Emittance {{Diagnostics}} at the {{MAX
  IV}} 3 {{GeV Storage Ring}}},
  \href{https://doi.org/10.18429/JACOW-IPAC2016-WEPOW034}{\emph{Proceedings of
  the 7th Int. Particle Accelerator Conf.} {\bfseries IPAC2016} (2016) 3 pages,
  0.811 MB}.

\bibitem{picoquant_notitle_nodate}
{PicoQuant}.
\newblock \href{http://www.picoquant.com}{http://www.picoquant.com}.

\bibitem{picoquant_pma}
{PicoQuant - PMA}.
\newblock
  \href{https://www.picoquant.com/products/category/photon-counting-detectors/pma-series-photomultiplier-detector-assembly}{https://www.picoquant.com/products/category/photon-counting-detectors/pma-series-photomultiplier-detector-assembly}.

\bibitem{idquantique_id100}
{idQuantique - ID100}.
\newblock
  \href{https://www.idquantique.com/quantum-detection-systems/products/id100/}{https://www.idquantique.com/quantum-detection-systems/products/id100/}.

\bibitem{tosi_fast-gated_2011}
A.~Tosi, A.D.~Mora, F.~Zappa, A.~Gulinatti, D.~Contini, A.~Pifferi et~al.,
  \emph{Fast-gated single-photon counting technique widens dynamic range and
  speeds up acquisition time in time-resolved measurements},
  \href{https://doi.org/10.1364/OE.19.010735}{\emph{Optics Express} {\bfseries
  19} (2011) 10735}.

\bibitem{becker_tcspc_2005}
W.~Becker, \emph{{{TCSPC Performance}} of the id100-50 {{Detector}}},  Nov.,
  2005.
\newblock
  \href{https://marketing.idquantique.com/acton/attachment/11868/f-0035/1/-/-/-/-/ID100-50\%20TCSPC\%20Report.pdf}{https://marketing.idquantique.com/acton/attachment/11868/f-0035/1/-/-/-/-/ID100-50\%20TCSPC\%20Report.pdf}.

\bibitem{picoquant_pma_hybrid}
{PicoQuant - PMA Hybrid}.
\newblock
  \href{https://www.picoquant.com/products/category/photon-counting-detectors/pma-hybrid-series-hybrid-photomultiplier-detector-assembly\#specification}{https://www.picoquant.com/products/category/photon-counting-detectors/pma-hybrid-series-hybrid-photomultiplier-detector-assembly\#specification}.

\bibitem{picoquant_picoharp_2014}
{PicoQuant}, \emph{{{PicoHarp}} 300 - {{Stand-alone TCSPC Module}}},  2014.
\newblock
  \href{http://www.picoquant.com/products/category/tcspc-and-time-tagging-modules/picoharp-300-stand-alone-tcspc-module-with-usb-interface}{http://www.picoquant.com/products/category/tcspc-and-time-tagging-modules/picoharp-300-stand-alone-tcspc-module-with-usb-interface}.

\bibitem{picoquant_picoharp_2024}
{PicoQuant}, \emph{{{PicoHarp}} 330 - {{Precise}} and {{Versatile Event Timer}}
  \& {{TCSPC Unit}}},  2024.
\newblock
  \href{https://www.picoquant.com/products/category/tcspc-and-time-tagging-modules/picoharp\_330\_precise\_and\_versatile\_event\_timer\_and\_tcspc\_unit}{https://www.picoquant.com/products/category/tcspc-and-time-tagging-modules/picoharp\_330\_precise\_and\_versatile\_event\_timer\_and\_tcspc\_unit}.

\bibitem{patting_dead-time_2007}
M.~Patting, M.~Wahl, P.~Kapusta and R.~Erdmann, \emph{Dead-time effects in
  {{TCSPC}} data analysis},
  \href{https://doi.org/10.1117/12.722804}{\emph{Proceedings of SPIE - The
  International Society for Optical Engineering} {\bfseries 6583} (2007)
  658307}.

\bibitem{brosi_studies_2017}
M.~Brosi, E.~Blomley, E.~Br{\"u}ndermann, M.~Caselle, B.~Kehrer, A.~Kopmann
  et~al., \emph{Studies of the {{Micro-Bunching Instability}} in {{Multi-Bunch
  Operation}} at the {{ANKA Storage Ring}}},  in \emph{Proc. of {{International
  Particle Accelerator Conference}} ({{IPAC}}'17), {{Copenhagen}}, {{Denmark}},
  14-19 {{May}}, 2017}, JACoW, May, 2017,
  \href{https://doi.org/10.18429/JACoW-IPAC2017-THOBA1}{DOI}.

\bibitem{tango_controls}
{Tango Controls}.
\newblock
  \href{https://www.tango-controls.org/}{https://www.tango-controls.org/}.

\bibitem{blanch-torne_control_nodate}
S.~{Blanch-Torn{\'e}}, ``Control {{Software}} for the {{PicoHarp300}}.''
\newblock
  \href{https://gitlab.com/srgblnch-tangocs/PicoHarp300}{https://gitlab.com/srgblnch-tangocs/PicoHarp300}.

\bibitem{skripka_commissioning_2016}
G.~Skripka, {\AA}.~Andersson, F.~Cullinan, A.~Mitrovic, R.~Nagaoka and
  P.~Tavares, \emph{Commissioning of the {{Harmonic Cavities}} in the {{MAX
  IV}} 3 {{GeV Ring}}},
  \href{https://doi.org/10.18429/JACOW-IPAC2016-WEPOW035}{\emph{Proceedings of
  the 7th Int. Particle Accelerator Conf.} {\bfseries IPAC2016} (2016) 3 pages,
  1.149 MB}.

\bibitem{hamamatsu_-__streak_camera_c10910_notitle_nodate}
{Hamamatsu - Streak Camera C10910}.
\newblock
  \href{https://www.hamamatsu.com/eu/en/product/photometry-systems/streak-camera/universal-streak-camera/C10910-05.html}{https://www.hamamatsu.com/eu/en/product/photometry-systems/streak-camera/universal-streak-camera/C10910-05.html}.

\bibitem{scikit-image-deconvolution}
{scikit-image}, ``Scikit-image deconvolution with the {{Richardson-Lucy}}
  algorithm.''
\newblock
  \href{https://scikit-image.org/docs/stable/auto\_examples/filters/plot\_deconvolution.html}{https://scikit-image.org/docs/stable/auto\_examples/filters/plot\_deconvolution.html}.

\bibitem{richardson_bayesian-based_1972}
W.H.~Richardson, \emph{Bayesian-{{Based Iterative Method}} of {{Image
  Restoration}}}, \href{https://doi.org/10.1364/JOSA.62.000055}{\emph{Journal
  of the Optical Society of America} {\bfseries 62} (1972) 55}.

\bibitem{ng_physics_2006}
K.Y.~Ng, \emph{Physics of Intensity Dependent Beam Instabilities}, World
  Scientific, Hoboken, NJ (2006),
  \href{https://doi.org/10.1142/5835}{10.1142/5835},
  \href{https://doi.org/10.1142/5835}{DOI}.

\bibitem{brosi_time-resolved_2023}
M.~Brosi, F.~Cullinan, A.~Andersson, J.~Breunlin and P.~Tavares,
  \emph{Time-resolved measurement and simulation of a longitudinal single-bunch
  instability at the {{MAX IV}} 3 {{GeV}} ring},  in \emph{Proceedings of the
  5th {{Int}}. {{Particle Accelerator Conf}}.}, pp.~2685--2688 pages, 1.2 MB,
  JACoW Publishing, 26.09.23,
  \href{https://doi.org/10.18429/JACOW-IPAC2023-WEPA020}{DOI}.

\bibitem{chao_physics_1993}
A.W.~Chao, \emph{Physics of Collective Beam Instabilities in High-Energy
  Accelerators} (1993),
  \href{https://www.slac.stanford.edu/{\textasciitilde}achao/wileybook.html}{https://www.slac.stanford.edu/{\textasciitilde}achao/wileybook.html}.

\end{thebibliography}\endgroup

%
%
%
%
%
\end{document}